\documentclass[a4paper,preprintnumbers,floatfix,superscriptaddress,pra,twocolumn]{revtex4-1}

\usepackage{amsmath, amsthm, amssymb}
\usepackage{graphicx}
\usepackage{dcolumn}
\usepackage{bm}
\usepackage{bbm}
\usepackage{color}
\usepackage[usenames,dvipsnames]{xcolor}
\usepackage{hyperref}
\usepackage{enumerate}
\usepackage{ulem}
\usepackage{textcomp}
\usepackage{multirow}

\newcommand{\figref}[1]{Fig.~\ref{#1}}

\newcommand{\secref}[1]{Sec.~\ref{#1}}

\newcommand{\ket}[1]{| #1 \rangle}
\newcommand{\bra}[1]{\langle #1 |}

\renewcommand{\t}[1]{\textrm{#1}}
\newcommand{\Tr}{\operatorname{Tr}}
\newcommand{\Hmin}{\t{H}_{\t{min}}}

\newcommand{\ba}{\begin{eqnarray}}
\newcommand{\ea}{\end{eqnarray}}

\begin{document}

\title{A self-testing quantum random number generator}

\author{Tommaso Lunghi}\thanks{These authors contributed equally to this work.}\affiliation{Group of Applied Physics, Universit\'e de Gen\`eve, 1211 Gen\`eve, Switzerland}
\author{Jonatan Bohr Brask}\thanks{These authors contributed equally to this work.}\affiliation{D\'epartement de Physique Th\'eorique, Universit\'e de Gen\`eve, 1211 Gen\`eve, Switzerland}
\author{Charles Ci Wen Lim}\affiliation{Group of Applied Physics, Universit\'e de Gen\`eve, 1211 Gen\`eve, Switzerland}
\author{Quentin Lavigne}\affiliation{Group of Applied Physics, Universit\'e de Gen\`eve, 1211 Gen\`eve, Switzerland}
\author{Joseph Bowles}\affiliation{D\'epartement de Physique Th\'eorique, Universit\'e de Gen\`eve, 1211 Gen\`eve, Switzerland}
\author{Anthony Martin}\affiliation{Group of Applied Physics, Universit\'e de Gen\`eve, 1211 Gen\`eve, Switzerland}
\author{Hugo Zbinden}\affiliation{Group of Applied Physics, Universit\'e de Gen\`eve, 1211 Gen\`eve, Switzerland}
\author{Nicolas Brunner}\affiliation{D\'epartement de Physique Th\'eorique, Universit\'e de Gen\`eve, 1211 Gen\`eve, Switzerland}

\begin{abstract}
The generation of random numbers is a task of paramount importance in modern science. A central problem for both classical and quantum randomness generation is to estimate the entropy of the data generated by a given device. Here we present a protocol for self-testing quantum random number generation, in which the user can monitor the entropy in real-time. Based on a few general assumptions, our protocol guarantees continuous generation of high quality randomness, without the need for a detailed characterization of the devices. Using a fully optical setup, we implement our protocol and illustrate its self-testing capacity. Our work thus provides a practical approach to quantum randomness generation in a scenario of trusted but error-prone devices.
\end{abstract}

\maketitle


Given the importance of randomness in modern science and beyond, e.g. for simulation algorithms and for cryptography, an intense research effort has been devoted to the problem of extracting randomness from quantum systems. Devices for quantum random number generation (QRNG) are now commercially available. All these schemes work essentially according to the same principle, exploiting the randomness of quantum measurements. A simple realization consists in sending a single photon on a 50/50 beam-splitter and detecting the output path \cite{rarity1994,stefanov2000,jennewein2000}. Other designs were developed, based on measuring the arrival time of single photons \cite{dynes2008,wahl2011,nie2014,stipcevic2007}, the phase noise of a laser \cite{qi2010,uchida2008,abellan2014}, vacuum fluctuations \cite{gabriel2010,symul2011}, and even mobile phone cameras \cite{sanguinetti2014}. 

A central issue in randomness generation is the problem of estimating the entropy of the bits that are generated by a device, i.e. how random is the raw output data. When a good estimate is available, appropriate post-processing can be applied to extract true random bits from the raw data (via a classical procedure termed randomness extractor \cite{nisan1999}). However, poor entropy estimation is one of the main weaknesses of classical RNG \cite{dodis2013}, and can have important consequences. In the context of QRNG, entropy estimates for specific setups were recently provided using sophisticated theoretical models \cite{frauchiger2013,ma2013}. Nevertheless, this approach has several drawbacks. First, these techniques are relatively cumbersome, requiring estimates for numerous experimental parameters which may be difficult to precisely assess in practice. Second, each study applies to a specific experimental setup, and cannot be used for other implementations. Finally, it offers no real-time monitoring of the quality of the RNG process, hence no protection against unnoticed misalignment (or even failures) of the experimental setup. 

It is therefore highly desirable to design QRNG techniques which can provide a real-time estimate of the output entropy. An elegant solution is provided by the concept of device-independent QRNG \cite{colbeckPhD,pironio2010}, where randomness can be certified and quantified without relying on a detailed knowledge of the functioning of the devices used in the protocol. Nevertheless, the practical implementation of such protocols is extremely challenging as it requires the genuine violation of Bell's inequality \cite{pironio2010,christensen2013}. Alternative approaches were proposed \cite{li2011,*li2012} but their experimental implementation suffers from loopholes \cite{dallArno2012}. More recently, an approach based on the uncertainty principle was proposed but requires a fully characterized measurement device \cite{vallone2014}.

Here, we present a simple and practical protocol for self-testing QRNG. Based on a prepare-and-measure setup, our protocol provides a continuous estimate of the output entropy. Our approach requires only a few general assumptions about the devices (such as quantum systems of bounded dimension) without relying on a detailed model of their functioning. This setting is relevant to real-world implementations of randomness generation, and is well-adapted to a scenario of trusted but error-prone providers, i.e.~a setting where the devices used in the protocol are not actively designed to fool the user, but where implementation may be imperfect. The key idea behind our protocol is to certify randomness from a pair of incompatible quantum measurements. As the incompatibility of the measurements can be directly quantified from experimental data, our protocol is self-testing. That is, the amount genuine quantum randomness can be quantified directly from the data, and can be separated from other sources of randomness such as fluctuations due to technical imperfections. We implemented this scheme with standard technology, using a single photon source and fibered telecommunication components. We implement the complete QRNG protocol, achieving a rate 23 certified random bits per second, with $99 \% $ confidence.

\begin{figure}[tbp]
\centering
\includegraphics[width=\linewidth]{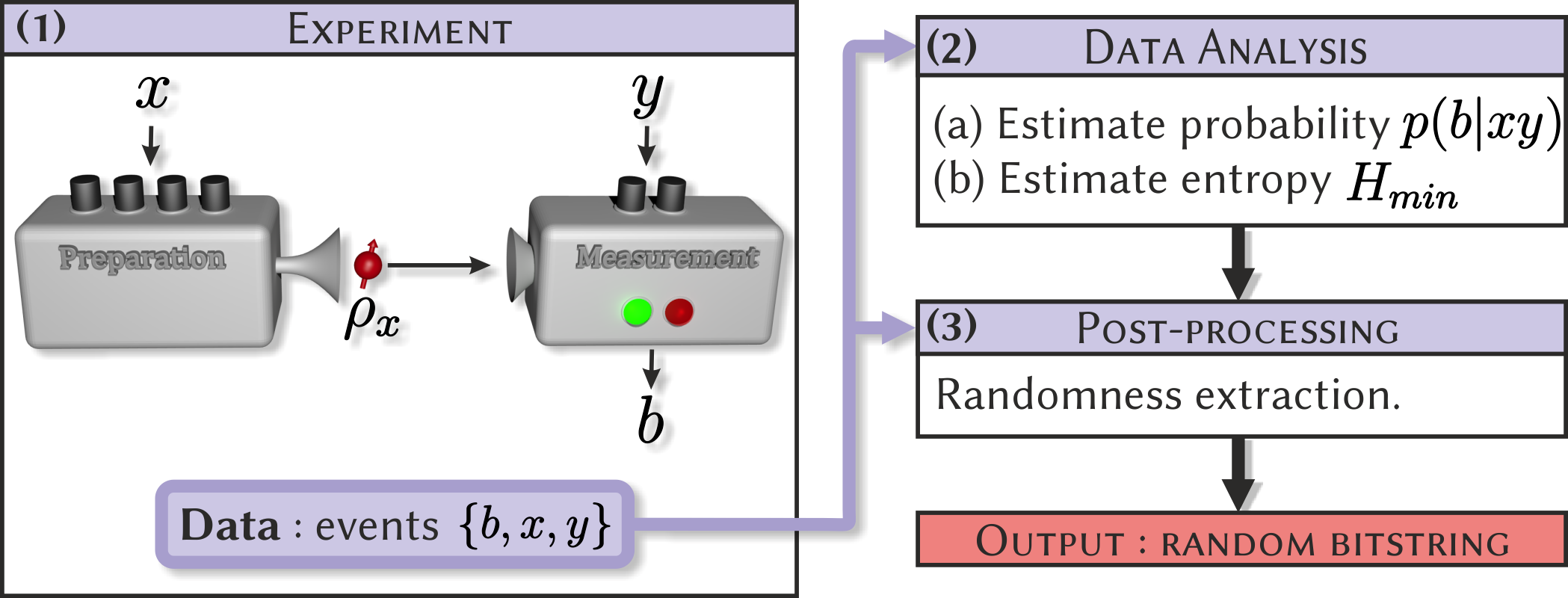}
\caption{\textbf{Sketch of the protocol.} The self-testing QRNG protocol consists in 3 distinct steps. \textbf{(1)} First, an experiment is performed where, in each round, the user chooses a preparation $x$ and a measurement $y$, and obtains an outcome $b$. \textbf{(2)} From the raw data, the distribution $p(b|x,y)$ can be estimated leading to an estimate for the value of the witness $W$, from which the entropy of the raw data can be quantified. \textbf{(3)} Based on the entropy bound, appropriate post-processing of the raw data is performed, in order to extract the final random bit string.}
\label{fig.model}
\end{figure}

{\it Protocol.} Our protocol, sketched in \figref{fig.model}, uses two devices which respectively prepare and measure an uncharacterized qubit system. In each round of the protocol, the observer chooses settings among four possible preparations, $x=0,1,2,3$, and two measurements $y=0,1$, resulting in a binary outcome $b=\pm1$. To model imperfections, we represent the internal state of each device by a random variable---$\lambda$ for the preparation device and $\mu$ for the measurement device---which are unknown to the observer. As we work in a scenario where the devices are not maliciously conspiring against the user, we assume the devices to be independent, \textit{i.e.} $p(\lambda,\mu)=q(\lambda)r(\mu)$, where $\int d\lambda q(\lambda)= \int d\mu r(\mu)=1$.

In each round of the experiment, the preparation device emits a qubit state $\rho_x^\lambda$ which depends on the setting $x$ and on the internal state $\lambda$. Similarly, the measurement device performs a measurement $M_y^\mu$. Thus the distributions of $\lambda$ and $\mu$ determine the distributions of the prepared states and the measurements. As the observer has no access to the variables $\lambda$ and $\mu$, he will observe
\ba1
p(b|x,y) &=& \int d\lambda q(\lambda) \int  d\mu \,  r(\mu) p(b|x,y,\lambda,\mu)  \nonumber \\
&=&  \Tr(\rho_x  \frac{\openone+ b M_y}{2} ) = \frac{1}{2} \left( 1+ b \vec{S}_x \cdot \vec{T}_y \right) 
\ea
where 
\ba \rho_x &=&  \int d\lambda q(\lambda) \rho_x^{\lambda} = \frac{1}{2} \left( \openone+ \vec{S}_x \cdot \vec{\sigma} \right)  \\
 M_{y} &=&  \int d\mu r(\mu) M_{y}^{\mu}  = \vec{T}_y \cdot \vec{\sigma}.
\ea
Here, $\vec{S}_x$ and $\vec{T}_y $ denote the Bloch vectors of the (average) states and measurements, and $\vec{\sigma}=(\sigma_1,\sigma_2,\sigma_3)$ is the vector of Pauli matrices. 

The task of the observer is to estimate the amount of genuine quantum randomness generated in this setup, based only on the observed distribution $p(b|x,y) $. This is a nontrivial task as the apparent randomness of the distribution ($0< p(b|x,y)< 1 $) can have different origins. On the one hand, it could be genuine quantum randomness. That is, if in a given round of the experiment, the state $\rho_x^\lambda$ is not an eigenstate of the measurement operator $M_{y}^\mu$, then the outcome $b$ cannot be predicted with certainty, even if the internal states $\lambda$ and $\mu$ are known, i.e. $0 < p(b|x,y,\lambda,\mu)< 1$. On the other hand, the apparent randomness may be due to technical imperfections, that is, to fluctuations of the internal states $\lambda$ and $\mu$. Consider the following example: The preparation device emits the states $\rho_{x}^{\lambda=0}=\ket{0}\bra{0}$ and $\rho_{x}^{\lambda=1}=\ket{1}\bra{1}$ with $q(\lambda=0,1)=1/2$. For a measurement of the observable $M_{y} =  \hat{z} \cdot \vec{\sigma} $, one obtains that $p(b|x,y)=1/2$. However, this data clearly contains no quantum randomness, since the outcome $b$ can be perfectly guessed if the internal state $\lambda$ is known. 

Our protocol allows the observer to separate quantum randomness from the randomness due to technical noise. The key technical tool of our protocol is a function recently presented in \cite{bowles2013}, which works as a 'dimension witness'. Given data $p(b|x,y)$, the quantity 
\begin{equation}
\label{eq.defW}
W=\left|
\begin{matrix}
p(1|0,0)-p(1|1,0) && p(1|2,0)-p(1|3,0) \\
p(1|0,1)-p(1|1,1) && p(1|2,1)-p(1|3,1)
\end{matrix}\right| .
\end{equation}
captures the quantumness of the preparation and measurements. Specifically, if the preparations are classical (i.e. there exist a basis in which all states $\rho_x^\lambda$ are diagonal), one has that $W=0$, while a generic qubit strategy achieves $0 \leq W \leq 1$ \cite{bowles2013}. $W>0$ guarantees that the measurements performed by Bob are incompatible (see \cite{suppinfo}) and since it is then impossible to simultaneously assign deterministic outcomes to them, this enables us to bound the guessing probability and certify randomness. Given $x,y$, and knowledge of the internal states $\lambda$, $\mu$, the best guess for $b$ is given by $\max_b p(b|x,y,\lambda,\mu) $. Assuming uniformly distributed $x$ and $y$, the average probability of guessing $b$ fulfils the following inequality (see \cite{suppinfo})
\begin{align}
\label{eq.pgbound}
p_{\text{guess}} & = \frac{1}{8} \sum_{x,y,\lambda,\mu} q_\lambda r_\mu \max_b p(b|x,y,\lambda,\mu) \nonumber \\
& \leq \frac{1}{2}\bigg(1+\sqrt{\frac{1+\sqrt{1-W^{2}}}{2}}\bigg).
\end{align}
Therefore the guessing probability can be upper-bounded by a function of $W$, which can be determined directly from the data $p(b|x,y)$. Finally, to extract random bits from the raw data, we use a randomness extraction procedure. The number of random bits that can be extracted per experimental run is given by the min-entropy $H_\text{min}= -\log_2 p_{\text{guess}}$ \cite{koenig2009}. Hence $H_\text{min}$ is the relevant parameter for determining how the raw data must be post-processed. Note that randomness can be extracted for any $W>0$, since $p_{\text{guess}}<1$ in this case.

The maximal value of $W=1$ can be reached using the set of preparations and measurements: $\vec{S}_0 = -\vec{S}_1 = \vec{T}_0 =   \hat{z}$ and $\vec{S}_2 = -\vec{S}_3 =  \vec{T}_1 =  \hat{x}$, which correspond to the BB84 QKD protocol~\cite{bennett1984}. In this case, we can certify randomness with min-entropy $H_\text{min} \simeq 0.2284 $. Using other preparations and measurements, \textit{e.g.} if the system is noisy or becomes misaligned, one will typically obtain $0<W<1$. Nevertheless, for any value $W>0$, randomness can be certified, and the corresponding min-entropy can be estimated using equation \eqref{eq.pgbound}. Our protocol is therefore self-testing, since the evaluation of $W$ allows quantifying the amount of randomness in the data. In turn, this allows one to perform adapted post-processing in order to finally extract random bits.

To conclude this section, we discuss the assumptions which are required in our protocol:
\begin{enumerate}[(i)]
\item \emph{Choice and distribution of settings.} The devices make no use of any prior information about the choice of settings $x$ and $y$.  
\item \emph{Internal states of the devices are independent and identically distributed (i.i.d).} The distributions $q(\lambda)$ and $r(\mu)$ do not vary between experimental rounds. 
\item \emph{Independent devices.} The preparation and measurement devices are independent, in the sense that $p(\lambda,\mu)=q(\lambda)r(\mu)$.
\item \emph{Qubit channel capacity.} The information about the choice of preparation $x$ retrieved by the measurement device (via a measurement on the mediating particle) is contained in a 2-dimensional quantum subspace (a qubit). 
\end{enumerate}

Assumptions \textrm{(i)} and \textrm{(iii)} are arguably rather natural in a setting where the devices are produced without malicious intent. They concern the independence of devices used in the protocol, namely the preparation and measurement devices, and the choice of settings. When these are produced by trusted (or simply different) providers, it is reasonable to assume that there are no (built-in) pre-established correlations between the devices and that the settings $x,y$ can be generated independently, \textit{e.g.} using a pseudo-RNG. Assumptions \textrm{(ii)} and \textrm{(iv)} are stronger, and will have to be justified for the particular implementation at hand. The content of assumption \textrm{(ii)} is essentially that the devices are memoryless (internal states do not depend on previous events). We believe this assumption can likely be weakened, since randomness can in fact be guaranteed in the presence of certain memory effects, in particular the experimentally relevant afterpulsing effect (see \cite{suppinfo}). Finally, note that assumption \textrm{(iv)} restricts the amount of information about $x$ that is retrieved by the measuring device (via a measurement on the mediating particle), but not the information about $x$ contained in the mediating particle itself. In other words, it might be the case that information about $x$ leaks out from the preparation device via side-channels, but we assume that these side-channels are not maliciously exploited by the measurement device.

\begin{figure*}[t!]
\centering
\includegraphics[width=\linewidth]{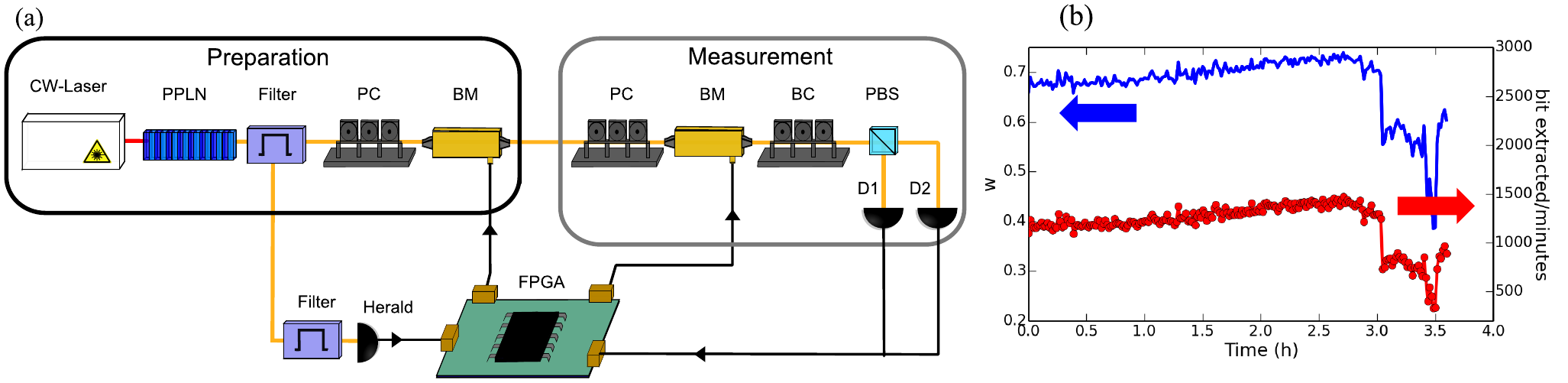}
\caption{Implementing the self-testing QRNG. (a) Experimental setup. (b) Real-time evolution of the witness value $W$ (blue) and randomness generation rate (bits extracted per second; red). After 3 hours, the air conditioning in the laboratory is switched off, which leads to misalignment of the optical components. In turn, this leads to a significant drop of the witness value $W$ and corresponding entropy.}
\label{fig.exp}
\end{figure*}

{\it Experiment.} We implemented the above protocol using a fully-guided optical setup (see \figref{fig.exp} (a)). The qubit preparations are encoded in the polarization state of single photons, generated via a heralded single-photon source based on a continuous wave spontaneous parametric down-conversion process in a periodically poled lithium niobate (PPLN) waveguide~\cite{tanzilli2012}. The idler photon is detected with a ID220 free-running InGaAs/InP single-photon detector (SPD) (herald) with 20\% detection efficiency and 20~\textmu s dead time. The polarization is rotated using a polarization controller (PC) and an electro-optical birefringence modulator (BM) based on a lithium niobate waveguide phase modulator. The preparations $x=\{ 0,1,2,3\}$ correspond respectively to the diagonal (D), anti-diagonal (A), circular right (R) and circular left (L) polarization states. For the measurement device, polarization measurements are done using a BM and a PC followed by a polarization beam splitter (PBS) and two ID210 InGaAs/InP SPDs (with a 1.5~ns gate and 25\% detection efficiency) triggered by a detection at the heralding detector. The measurements $y = \{0,1\}$ correspond respectively to the \{D,A\} basis and the \{R,L\} basis. The number of photon pairs generated by the SPDC source is set to obtain a count rate at the heralding detector of about $30$~kHz, which corresponds to a probability of single photon emission of $p_1=6.5\times 10^{-4}$ per gate, and a two photon emission $p_2 = p^2_1/2 = 2.1\times 10^{-7}$ per gate. A Field-Programmable-Gate-Array board (FPGA) continuously generates sequences of 3 pseudo-random bits. Upon successful heralding, these 3 bits are used to choose ($x,y$). Finally, the FPGA records the outcome $b$ (whether each ID210 detector has clicked or not).

We briefly discuss to which extent the assumptions of the protocol fit to our implementation. First, the choice of preparation and measurement, $x$ and $y$, are made by the FPGA using a linear-feedback shift register pseudo-RNG~\cite{PSEUDORNG}. This RNG provides a deterministic cyclic function sampled by the heralding detector. Since the sampling is asynchronous with respect to the RNG rate, the output is uniform and \textrm{(i)} is fulfilled. The BMs are separated spatially by 1\,m, their temperature is controlled independently, and the voltages are applied with independent electronic circuits. Any cross-talk between them, \textit{e.g.} due to stray electric fields, can be safely neglected, hence \textrm{(iii)} is also fulfilled. Concerning assumption \textrm{(ii)}, we evaluate the distribution $p(b|x,y)$ after every minute of acquisition. Therefore, we need to consider memory effects with time characteristics shorter than 1 minute. Two main effects should be considered: charge accumulation in the birifringence modulator, and afterpulsing in the detectors, which is a common issue in standard QRNG approaches \cite{dynes2008,frauchiger2013}. Importantly, our protocol is robust to afterpulsing, (see \cite{suppinfo}). Charge effects in the modulator are relevant only for modulation slower than 1~Hz \cite{Wooten2000}. Finally, the qubit assumption \textrm{(iv)} is arguably the most delicate one. As the choice of preparation $x$ is encoded in the polarization of a single photon, \textrm{(iv)} seems justified. However, a small fraction of heralded events corresponds to multi-photon pulses, in which \textrm{(iv)} is not valid. To take these events into account, we extend our theoretical analysis (see \cite{suppinfo}). We show that quantum randomness can still be guaranteed even when \textrm{(iv)} is not fulfilled in all experimental events, provided that the fraction of events violating \textrm{(iv)} can be bounded and is small enough compared to the total number of successful events. To verify this assumption, the probability of single and multi-photon pulses must be properly calibrated. For our single-photon source, the ratio of multi-photon events vs.~heralds is given by $\sim p_1/2 = 3.25\times 10^{-4}$, and our method can be applied.

We ran the experiment estimating $W$ for the data accumulated each minute. As discussed in \cite{suppinfo}, the estimation of $W$ considers finite-size effects and the size of the randomness extractor is determined based on the value of $W$ \cite{troyer2012,frauchiger2013}. In the best conditions, our setup generates about 402 bits/s of raw data (before the extractor). The witness corresponds to a value of $W=0.76$. After extraction, we get final random bits at a rate of 23 bits/s with a confidence of $99 \% $. Note that the confidence level is set when accounting for finite size effects; a higher confidence can be chosen at the expense of a lower rate. Note also that this rate is limited by the slow repetition rate of the experiment (limited by the dead time of the heralding detector) and by the losses in the optical implementation (channel transmission is $\sim 8 \%$; total efficiency $\sim 2 \%$). Fig.~\ref{fig.exp}(b) shows the estimated value of $W$ over 3.5 hours and the rate at which the final random bits are generated. To demonstrate the self-testing capacity of our protocol, we switched off the air conditioning in the room after 3 hours. This impacts the alignment of the setup. As can be seen from \figref{fig.exp}(b), the witness value $W$ drops, reflecting the fact that the distributions of internal states ($q(\lambda)$ and $r(\mu)$) changed. In turn, this forces us to perform more post-processing, resulting in a lower randomness generation rate. Nevertheless, the quality of the final random bits is still guaranteed. This shows that our setup can warrant the generation of high quality randomness, without active stabilization or precise modelling of the impact of the temperature increase.

The quality of the generated randomness can be assessed by checking for patterns and correlations in the extracted bits. We performed standard statistical test, as defined by NIST, and although not all tests could be performed due to the small size of the sample, all performed tests were successful (see \cite{suppinfo}). We stress that these tests do not constitute a proof of randomness (which is impossible), however failure to pass any of them would indicate the presence of correlations among the output bits.

Finally, we comment on the influence of losses. In the above analysis, we discarded inconclusive events in which the photon was not detected at the measuring device, although the emission of a single-photon was heralded by the source. Therefore, our analysis is subject to an additional assumption, namely that of fair-sampling, which we believe is rather natural in the case of non-malicious devices. Note however that this is not necessary strictly speaking, as our protocol is in principle robust to arbitrarily low detection efficiency \cite{bowles2013}. Performing the data analysis without the fair-sampling assumption (in which case the inconclusive events are attributed the outcome -1) we obtain witness values of $W \sim 1.5\times 10^{-4}$, corresponding to $H_{min} \sim 2.0\times 10^{-9}$. In this case, the rate for generating random bits drops considerably to $6\times 10^{-5}$ bits/s, but importantly does not vanish. Hence, our setup can be used to certify randomness without requiring the fair-sampling assumption. We note that even a small increase in efficiency would lead to a large improvement in rate. E.g.~an increase from our current $2\%$ to $10\%$ would already give $\sim 0.04$ bits/s while an overall efficiency of $50\%$ would be enough to reach $23$ bits/s without post-selection, equal to our current post-selected rate.

{\it Conclusion.} We have presented a protocol for self-testing QRNG, which allows for real-time monitoring of the entropy of the raw data. This allows adapting the randomness extraction procedure in order to continuously generate high quality random bits. Using a fully optical guided implementation, we have demonstrated that our protocol is practical and efficient, and illustrated its self-testing capacity. Our work thus provides an approach to QRNG, which can be viewed as intermediate between the standard (device-dependent) approach and the device-independent one.

Compared to the device-dependent approach, our protocol delivers a stronger form of security requiring less characterization of the physical implementation, at the price of a reduced rate compared to commercial QRNGs such as ID Quantique QUANTIS which reaches 4Mbits/s. A fully device-independent approach \cite{colbeckPhD,pironio2010}, on the other hand, offers even stronger security (in particular assumptions (ii)-(iv) can be relaxed, hence offering robustness to side-channels and memory effects), but its practical implementation is extremely challenging. Proof-of-principle experiments require state-of-the-art setups but could achieve only very low rates \cite{pironio2010,christensen2013}. Our approach arguably offers a weaker form of security, but can be implemented with standard technology. Our work considers a scenario of trusted but error-prone devices, which we believe to be relevant in practice. 

\emph{Note added} After submission of this work, several related works have appeared \cite{mitchell2015,canas2014,haw2014}.

\emph{Acknowledgements.} We thank Antonio Acin, Stefano Pironio, Valerio Scarani, and Eric Woodhead for discussions; Raphael Houlmann and Claudio Barreiro for technical support; Batelle and ID Quantique for providing the PPLN waveguide. We acknowledge financial support from the Swiss National Science Foundation (grant PP00P2\_138917 and QSIT), SEFRI (COST action MP1006) and the EU project SIQS.

\appendix

\section*{Supplementary material}

In this Supplementary material we provide a proof of randomness for our protocol along with the required assumptions in \secref{app.rndproof}. We show that our protocol is robust to detector afterpulsing in \secref{app.afterpulsing}. We show how to account for multi-photon events in \secref{app.imperf}, and we account for finite-size effects in \secref{app.secanal}. Finally, we discuss statistical tests applied to the output data.

\section{Proof of randomness}
\label{app.rndproof}

Here we provide a lower bound on the randomness in the observed output using the dimension witness of Ref.~\cite{bowles2013}. The devices are assumed to be independent, but each device features an internal source of randomness, represented by the variable $\lambda$ for Alice, and variable $\mu$ for Bob. Our goal is to upper bound the probability of guessing the output $b$ that one would have if $\lambda$ and $\mu$ were known, averaged over all inputs and values of the local random variables. Before proceeding with the proof, we first establish the setting in which we will work and state the assumptions made.

\subsection{Setting and assumptions}

 A priori, the probability of observing a certain output in a given round of the experiment could depend on everything that happened before, and later events could be correlated with the observation of a certain output. However, we will introduce several assumptions which ensure that we can speak about output probabilities without referring to specific rounds as well as the independence of the devices. Let us associate random variables $B_i$, $X_i$, $Y_i$, $\Lambda_i$, $M_i$ with the output, the inputs, and the internal variables in round $i$, and let us write $\vec{B_i}$ for the set of variables $B_1,...,B_i$ etc. Also, let us denote the probabilities for the random variables to take on specific values by lower case symbols, e.g. $p(x_i) = P(X_i=x_i)$ and $p(\vec{b}_i|\vec{x}_i,\vec{y}_i) = P(\vec{B}_i=\vec{b}_i|\vec{X}_i=\vec{x}_i,\vec{Y}_i=\vec{y}_i)$. 

Our first assumption is that \textit{all inputs are independent of each other and the devices}. Formally, $X_i$ is independent of $X_j$ for any $j\neq i$ and of $\vec{Y}_{i-1}$, $\vec{\Lambda}_{i-1}$, $\vec{M}_{i-1}$, and similarly for $Y_i$. Our second assumption is that \textit{the output in a given round depends only on the inputs in that round and the current state of the devices}. Formally, $B_i$ is conditionally independent of $\vec{B}_{i-1}$, $\vec{X}_{i-1}$, $\vec{Y}_{i-1}$, $\vec{\Lambda}_{i-1}$, and $\vec{M}_{i-1}$ given $X_i$, $Y_i$, $\Lambda_i$, and $M_i$. Our third assumption is that \textit{the devices do not record the outputs}. Formally, $\Lambda_i$ and $M_i$ are independent of $\vec{B}_{i-1}$. Under these assumptions, the probability for a certain string of outputs to occur factorises
\begin{equation}
p(\vec{b}_n|\vec{x}_n,\vec{y}_n,\vec{\lambda}_n,\vec{\mu}_n) = \prod_{i=1}^n p(b_i|x_i,y_i,\lambda_i,\mu_i) .
\end{equation}
This can be seen by repeated application of Bayes' rule. The probability to correctly guess the output string $\vec{b}_n$ knowing all the inputs and internal variables in an experiment with $n$ rounds is
\begin{align}
p^g_{\vec{x}_n\vec{y}_n\vec{\lambda}_n\vec{\mu}_n} & = \underset{\vec{b}_n}{\max} \, p(\vec{b}_n|\vec{x}_n,\vec{y}_n,\vec{\lambda}_n,\vec{\mu}_n) \nonumber \\
& = \prod_{i=1}^n \underset{b}{\max} \, p(b|x_i,y_i,\lambda_i,\mu_i) ,
\end{align}
and it follows that
\begin{align}
\log(p^g_{\vec{x}_n\vec{y}_n\vec{\lambda}_n\vec{\mu}_n}) & =  \sum_{i=1}^n \log( \underset{b}{\max} \, p(b|x_i,y_i,\lambda_i,\mu_i) ) \nonumber  \\
& \leq n \log( \frac{1}{n}\sum_{i=1}^n \underset{b}{\max} \, p(b|x_i,y_i,\lambda_i,\mu_i) ) .
\end{align}
We now assume that \textit{the distribution of the internal randomness is fixed for the duration of the experiment}. Formally, the $\vec{\Lambda}_n$ are identically distributed, and the $\vec{M}_n$ as well. With this assumption, for $n \rightarrow \infty$ the sum in the last line above is equivalent to averaging over the inputs and internal variables, that is, it equals
\begin{equation}
\sum_{x,y}\sum_{\lambda,\mu} \underset{b}{\max} \, p(b|x,y,\lambda,\mu) p(x,y) p(\lambda,\mu) .
\end{equation}
With the final assumption that \textit{the devices are independent}, formally that the $\vec{\Lambda}_n$ are independent of the $\vec{M}_n$, it follows from our proof below that this quantity is bounded by a function of the observed witness value $f(W)$. This implies that in the limit of large $n$
\begin{equation}
p^g_{\vec{x}_n\vec{y}_n\vec{\lambda}_n\vec{\mu}_n} \leq f(W)^n ,
\end{equation}
and hence the entropy per bit in the output string is bounded by
\begin{equation}
H = -\frac{1}{n} \log_2(p^g_{\vec{x}_n\vec{y}_n\vec{\lambda}_n\vec{\mu}_n}) \geq -\log_2(f(W)) .
\end{equation}

We have assumed that the internal random variables are identically distributed in every round. On the physical level, the corresponding requirement is that any external parameters which influence the distributions $q_\lambda$, $r_\mu$, such as e.g.~temperature, vary slowly on the time-scale of one experimental run, i.e.~the time required to gather enough data to estimate the witness value $W$. In our experimental implementation this time-scale is about one minute. Between different experimental runs there is no requirement for $q_\lambda$, $r_\mu$ to stay unchanged. We have also assumed that the internal variables are independent of the outputs. Note however that we believe that these assumptions can be relaxed. For example, detector afterpulsing breaks the second assumption, but randomness can nevertheless be certified in our protocol as demonstrated in \secref{app.afterpulsing}.

\subsection{Proof}

Having established the above assumptions, we can now go ahead with our randomness proof without reference to any specific round of the experiment, i.e.~we can work just with the distribution $p(b|x,y,\lambda,\mu)$. For given inputs and $\lambda$, $\mu$, the guessing probability for this distribution is
\begin{equation}
p^g_{xy\lambda\mu} = \max_b p(b|x,y,\lambda,\mu) .
\end{equation}
The average guessing probability $p^g$ is the average of $p^g_{xy\lambda\mu}$ over the distribution of inputs and local randomness. To proceed, however, we will first derive an upper bound on $p^g_{\lambda\mu}$, defined to be the average over the inputs only.

\begin{figure}[t]
\centering
\includegraphics[width=0.85\linewidth]{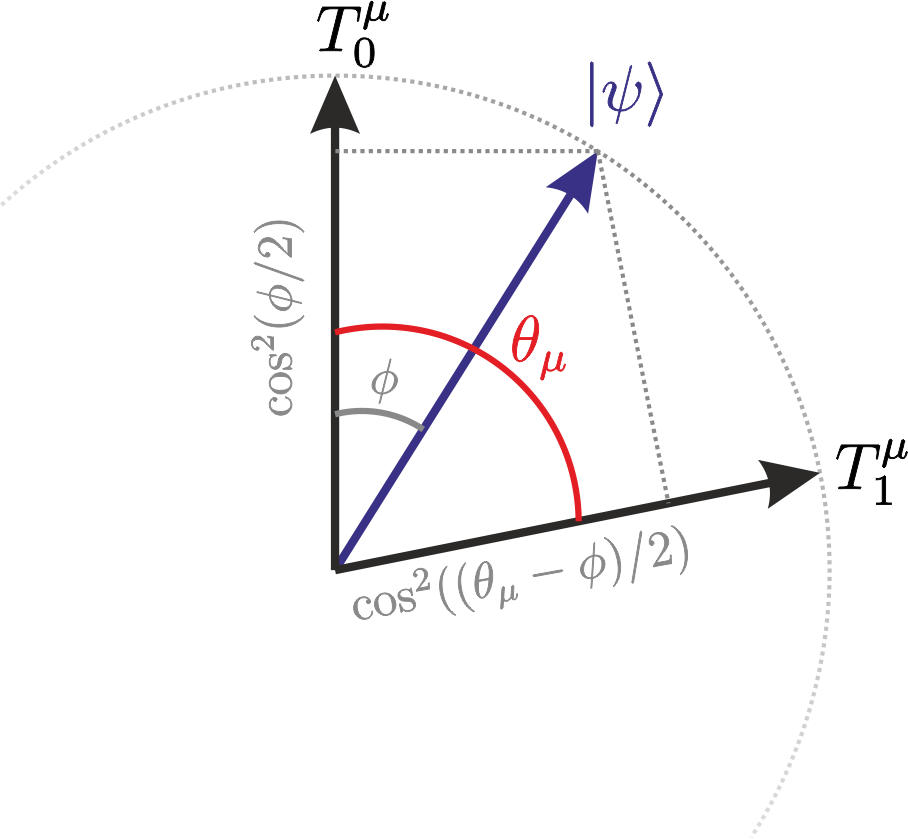}
\caption{Cut through the Bloch sphere showing the measurements of Bob, and a state $\ket{\psi}$ lying in the same plane. The probabilities of outcome, say, $b=1$ are given by the projections of $\ket{\psi}$ onto $T^\mu_{0,1}$. The probabilities when $\ket{\psi}$ makes an angle $\phi$ with $T^\mu_0$ are indicated. To maximise the average of these, one must choose $\phi = \theta_\mu/2$. Note that choosing a state out of the plane of the measurements can only decrease the guessing probability.}
\label{fig.measgeom}
\end{figure}

We consider the witness $W$ of the main text. We thus have four preparations, $x=0,1,2,3$ and two measurements $y=0,1$. Consider choices of preparations and measurements which are uniformly random (as explained in the main text, pseudorandomness is sufficient here), i.e. each combination $x,y$ occurs with probability $1/8$. We have that
\begin{equation}
\label{eq.pglambdamubound}
\begin{split}
p^g_{\lambda\mu} & = \frac{1}{8} \sum_{x,y} \max_b p(b|x,y,\lambda,\mu) \\
& \leq \frac{1}{2} \max_x \sum_y \max_b p(b|x,y,\lambda,\mu) \\
& \leq \frac{1+ \cos(\theta_\mu/2)}{2}
\end{split}
\end{equation}
where $\theta_\mu$ denotes the angle between Bob's two measurement. The reasoning of the derivation is as follows. The best guessing probability averaged over inputs of Alice is bounded by the maximum over her inputs. This gives the first inequality and allows us to focus on the best possible state that Alice can send. Next, Bob has two measurements described by Bloch vectors $\vec{T}_{0,1}^\mu$, and $\theta_\mu$ is the angle between them. The best guessing probability averaged over his inputs is obtained by sending a state which lies in the middle between his measurements on the Bloch sphere (see \figref{fig.measgeom}). For such a state, the outcome probabilities for the two values of $b$ are $\cos^2(\theta_\mu/4)$, and $\sin^2(\theta_\mu/4)$. Choosing the larger value and using the double-angle formula, one arrives at the second inequality.

Now we use the fact that a bound on the angle $\theta_\mu$ can be derived from the witness value for fixed local randomness $W_{\lambda,\mu}$. One has that (see \cite{bowles2013})
\begin{eqnarray}
\label{eq.Wlambdamubound}
W_{\lambda,\mu} \leq |\vec{T}_{0}^\mu \times \vec{T}_{1}^\mu| \leq \sin{\theta_\mu}
\end{eqnarray}
For maximally anti-commuting measurements, we get $W_{\lambda,\mu}=1$. Combining \eqref{eq.pglambdamubound} and \eqref{eq.Wlambdamubound}, we get 
\begin{eqnarray}
\label{eq.pghiddenbound}
p^g_{\lambda,\mu}  \leq \frac{1}{2} \left( 1+  \sqrt{ \frac{1+\sqrt{1-W^2_{\lambda,\mu} }}{2}  } \right)  \equiv f(W_{\lambda,\mu}) .
\end{eqnarray}
We note that the function $f$ is concave and decreasing.

Next, we establish the following convexity property of the witness (in a slight abuse of notation, $W$ denotes the observed value of the witness when $\lambda$, $\mu$ are not known)
\begin{eqnarray}
\label{eq.Wconvex}
W \leq \sum_{\lambda,\mu} q_\lambda r_\mu W_{\lambda,\mu} .
\end{eqnarray}
To see that this holds, consider the entries of the matrix defining $W$. They are of the form $p(1|x,y) - p(1|x',y)$. When the devices have internal randomness, we can write
\begin{align}
p(1|x,y)-p(1|x',y) & = \sum_{\lambda,\mu} q_\lambda r_\mu \left( \Tr [\rho_x^\lambda \Pi_{1|y}^\mu ] - \Tr [\rho_{x'}^\lambda \Pi_{1|y}^\mu ]\right) \nonumber \\
& =  \sum_{\lambda,\mu} q_\lambda r_\mu \vec{S}_{xx'}^\lambda \cdot \vec{T}_y^\mu \nonumber \\
& = \left( \sum_\lambda q_\lambda \vec{S}_{xx'}^\lambda \right) \cdot \left( \sum_\mu r_\mu \vec{T}_y^\mu \right) \nonumber \\
& \equiv \vec{S}_{xx'} \cdot \vec{T}_y ,
\end{align}
where $\rho_x^\lambda$ are the states produced by Alice's box, and $\Pi_{1|y}^\mu = (\mathbbm{1} + M^\mu_y)/2$ are the projection operators of Bob corresponding to outcome 1, $\vec{T}_y^\mu$ is the Bloch vector corresponding to $M_y^\mu$ and $S_{xx'}^\lambda$ is the difference of the Bloch vectors for $\rho_x^\lambda$ and $\rho_{x'}^\lambda$ (see \cite{bowles2013}). Now, from \cite{bowles2013} it follows that
\begin{align}
W & = (S_{01} \times S_{23})\cdot(T_0 \times T_1) \\
& = \sum_{\lambda, \lambda', \mu, \mu'} q_\lambda q_{\lambda'} r_\mu r_{\mu'} (S_{01}^\lambda \times S_{23}^{\lambda'}) \cdot (T_0^\mu \times T_1^{\mu'}) \\
& = \sum_{\lambda, \lambda', \mu, \mu'} q_\lambda q_{\lambda'} r_\mu r_{\mu'} |S_{01}^\lambda \times S_{23}^{\lambda'}| |T_0^\mu \times T_1^{\mu'}| \cos \phi_{\lambda,\lambda',\mu,\mu'} 
\end{align}
where $\phi_{\lambda,\lambda',\mu,\mu'} $ denotes the angle between the vectors $(S_{01}^\lambda \times S_{23}^{\lambda'})$ and $(T_0^\mu \times T_1^{\mu'})$. Next we notice that, for fixed $\lambda,\mu,\mu'$, there will be a value of $\lambda'$ such that $|S_{01}^\lambda \times S_{23}^{\lambda'}| \cos \phi_{\lambda,\lambda',\mu,\mu'}$ is maximal. If we label this value $\lambda$ and set $q_{\lambda'} = 1$ when $\lambda'=\lambda$ this can only increase the expression. We thus obtain:

\begin{align}
W & \leq \sum_{\lambda, \mu, \mu'} q_\lambda r_\mu r_{\mu'} |S_{01}^\lambda \times S_{23}^\lambda| |T_0^\mu \times T_1^{\mu'}| \cos \phi_{\lambda,\mu,\mu'} 
\end{align}
Using a similar argument, we can eliminate $\mu'$:
\begin{align}
W & \leq \sum_{\lambda, \mu} q_\lambda r_\mu |S_{01}^\lambda \times S_{23}^\lambda| |T_0^\mu \times T_1^\mu| \cos \phi_{\lambda,\mu} \\
& = \sum_{\lambda, \mu} q_\lambda r_\mu W_{\lambda,\mu} .
\end{align}
We are now ready to bound the guessing probability $p^g$. Using the definition of $p^g$, \eqref{eq.pghiddenbound}, and \eqref{eq.Wconvex} we have
\begin{align}
p^g & = \sum_{\lambda,\mu} q_\lambda r_\mu p^g_{\lambda,\mu} \\
& \leq \sum_{\lambda,\mu} q_\lambda r_\mu f(W_{\lambda,\mu}) \\
& \leq  f(\sum_{\lambda,\mu} q_\lambda r_\mu W_{\lambda,\mu}) \\
& \leq f(W)
\end{align}
where in the third line we have used Jensen's inequality and concavity of $f$, and in the last line we have used that $f$ is decreasing. Hence, we finally get \begin{eqnarray} p^g \leq \frac{1}{2} \left( 1+  \sqrt{ \frac{1+\sqrt{1-W^2 }}{2}  } \right)  \end{eqnarray}
which gives the desired upper bound on the guessing probability as a function of the observed value of the witness $W$. This bound is tight when maximal violation of the witness is achieved, i.e. $W=1$. In \secref{app.secanal}, we provide the calculation for the maximum number of extractable random bits.

Finally, we provide a proof of the relation between $W$ and the commutativity of the measurements. We write $M^\mu_y = \vec{T}^\mu_y \cdot \vec{\sigma}$, and we have
\begin{align}
\int d \mu r(\mu)  ||\left[  M_0^\mu , M_1^\mu \right] || & = \int d \mu r(\mu)  ||\left[  \vec{T}^\mu_0 \cdot \vec{\sigma} , \vec{T}^\mu_1 \cdot \vec{\sigma} \right] || \nonumber \\
& =  \int d \mu r(\mu)  || 2i (\vec{T}^\mu_0 \times \vec{T}^\mu_1) \cdot \vec{\sigma} || \nonumber \\
& = 2 \int d \mu r(\mu)  |\vec{T}^\mu_0 \times \vec{T}^\mu_1| \nonumber \\
& \geq 2 \int d\lambda d\mu q(\lambda)r(\mu) W_{\lambda,\mu} \nonumber \\
& \geq 2 W ,
\end{align}
where we have used \eqref{eq.Wlambdamubound} and \eqref{eq.Wconvex}.

\section{Certifying randomness in the presence of afterpulsing}
\label{app.afterpulsing}

In the following we show that although afterpulsing a priory violates the i.i.d.~assumption \textit{(iii)}, the self-testing nature of our protocol captures the effect. When afterpulsing is present, the witness value is reduced correspondingly and randomness can still be certified.

To see this, we first consider a hypothetical experiment in which the outputs are generated as follows: in a fraction $\eta$ of events, the experiment follows and ideal quantum qubit implementation while for the remaining events an outcome is generated at random by the measurement device, determined only by some internal random variable $\mu$ independent of the inputs. Let us denote the witness value computed from the whole dataset $W$, and the value which would be obtained from only the quantum events $\tilde{W}$. To an observer who does not know $\mu$, the non-quantum events look just like uniform noise and the witness values fulfil $W = \eta^2 \tilde{W}$ \cite{bowles2013}. At the same time, this scenario meets all of the assumptions in the proof of randomness of \secref{app.rndproof}. Therefore, for an observer with perfect knowledge of $\mu$, who can hence perfectly predict the output for the non-quantum events, the guessing probability on the whole dataset is bounded by
\begin{equation}
\label{eq.pgboundmixture}
p^g \leq f(W) = f(\eta^2 \tilde{W}) .
\end{equation}
We now show that the witness value is reduced in a similar way for afterpulsing, and hence even if the outputs from afterpulsing events can be perfectly predicted, our bound on the randomness still holds. 

Consider an experiment generating a set $S=\{(b_1,x_1,y_1),\ldots,(b_N,x_N,y_N)\}$ of $N$ events. The first thing to notice is that afterpulsing is probabilistic: in any given event either there is an afterpulse or there is not. We can therefore think of $S$ as consisting of a set $\tilde{S}$ of $\tilde{N}$ events with no afterpulse and $N-\tilde{N}$ additional afterpulsing events. Let $N_{bxy}$ denote the number of events in $S$ with outcome $b$ and inputs $x$, $y$, and $\tilde{N}_{bxy}$ the events in $\tilde{S}$, and define $N_{xy}$, $\tilde{N}_{xy}$ similarly. For simplicity let us consider the limit of large $N$ such that finite size effects can be neglected. Since the inputs are chosen uniformly $N_{xy} = N/8$. We note that the probability for an afterpulse to occur in a given round $i$ of the experiment does not depend on the inputs $x_i$, $y_i$ in that round. The number of afterpulses is therefore the same for all combinations of $x$,$y$, and $\tilde{N}_{xy} = \eta N / 8$ with $\eta = \tilde{N}/N$. In any afterpulsing event, the outcome $b_i$ is also uncorrelated to the inputs $x_i$, $y_i$ in that round (since $b_i = b_{i-1}$). This means that the effect of afterpulsing when counting events can be written
\begin{equation}
N_{b,x,y} = \tilde{N}_{b,x,y} + c_b ,
\end{equation}
where, importantly, $c_b$ is independent of $x$ (also of $y$ and indeed it may be independent of $b$, but this is not important in the following).

The witness value on the dataset $S$ is computed from the frequencies $\nu_{b|xy} = N_{b,x,y}/N_{x,y}$. Using the above, we can write
\begin{equation}
\nu_{b|xy} = \frac{\tilde{N}_{b,x,y}+c_b}{N/8} = \frac{\eta\tilde{N}_{b,x,y}}{\eta N/8} + \frac{8c_b}{N} = \eta\tilde{\nu}_{b|xy} + \frac{8c_b}{N} ,
\end{equation}
where $\tilde{\nu}_{b|xy} = \tilde{N}_{b,x,y}/\tilde{N}_{x,y}$ is the frequency one would have obtained considering only the set $\tilde{S}$. Now, since the last term above is independent of $x$ and since the witness is computed solely from terms of the form $\nu_{1|xy} - \nu_{1|x'y}$, we have that
\begin{equation}
W = \eta^2 \tilde{W} ,
\end{equation}
where $\tilde{W}$ is the witness value which one would obtain from the events $\tilde{S}$ without afterpulsing. Since the reduction in $W$ when afterpulses are added is exactly the same as in the scenario above where events with perfectly predictable outputs were added, it follows that even if afterpulse events would be perfectly predictable, the bound \eqref{eq.pgboundmixture} on the guessing probability still holds.

\section{Accounting for multi-photon events}
\label{app.imperf}

For real-world sources it is challenging to guarantee that they are of qubit nature. In particular, single-photon sources based on spontaneous parametric down conversion process or weak coherent sources have non-zero probability of emitting more than one photon, violating the qubit assumption.

Given an imperfect source which does not always satisfy the qubit assumption, we would like to say something about the witness violation corresponding to events that do satisfy the assumption. In particular, we would like a lower bound on this violation in terms of the observed, experimental probability distribution and some guarantee on the fraction of non-qubit events. Even without a detailed model of the source, it is possible to determine this fraction e.g.~using knowledge of the photon statistics.

\subsection{Bounding the violation for given qubit fraction}

To derive a bound on the quantum violation, we will assume that each experimental round either satisfies the qubit assumption, or not. That is, the conditional probability distribution for the experiment can be modeled as
\begin{equation}
\label{eq.pdecomp}
p(b|xz) = \alpha p_{qa}(b|xz) + (1-\alpha) p_{\bar{qa}}(b|xz) ,
\end{equation}
where $\alpha$ is the fraction of qubit events, $p_{qa}$ is the distribution corresponding to the qubit events, and $p_{\bar{qa}}$ is an unrestricted distribution. The witness value is given in terms of the probabilities by $|W|$, where
\begin{equation}
W=\left|
\begin{matrix}
p(1|0,0)-p(1|1,0) && p(1|2,0)-p(1|3,0) \\
p(1|0,1)-p(1|1,1) && p(1|2,1)-p(1|3,1)
\end{matrix}\right| .
\end{equation}
From the model \eqref{eq.pdecomp}, it follows that the expected witness value must satisfy
\begin{equation}
\label{eq.Wobs}
W = |\alpha^2 W_{qa} + (1-\alpha)^2 W_{\bar{qa}} + \alpha (1-\alpha) (G + G')| ,
\end{equation}
where $W_{qa}$, $W_{\bar{qa}}$ are the determinants corresponding to distributions $p_{qa}$ and $p_{\bar{qa}}$ respectively, and 
\begin{equation*}
G=
\left|\begin{matrix}
p_{qa}(1|0,0)-p_{qa}(1|1,0) && p_{qa}(1|2,0)-p_{qa}(1|3,0) \\
p_{\bar{qa}}(1|0,1)-p_{\bar{qa}}(1|1,1) && p_{\bar{qa}}(1|2,1)-p_{\bar{qa}}(1|3,1)
\end{matrix}\right|
\end{equation*}
\begin{equation*}
G'=
\left|\begin{matrix}
p_{\bar{qa}}(1|0,0)-p_{\bar{qa}}(1|1,0) && p_{\bar{qa}}(1|2,0)-p_{\bar{qa}}(1|3,0) \\
p_{qa}(1|0,1)-p_{qa}(1|1,1) && p_{qa}(1|2,1)-p_{qa}(1|3,1)
\end{matrix}\right|
\end{equation*}
To bound the qubit violation for a given expected observed violation we should minimise $|W_{qa}|$ subject to the constraint \eqref{eq.Wobs}. However, if a certain value $W$ can be attained for a fixed value of $|W_{qa}|$, then attaining all smaller values requires even less qubit violation. We may therefore just as well look for the maximal $W$ for fixed $|W_{qa}|$. Any value above this maximum guarantees a qubit violation of at least $|W_{qa}|$. The maximum has a simple form. It is given by 
\begin{equation}
\label{eq.maxWobs}
\max W = \max \left\{ \begin{matrix}
4\alpha(1-\alpha) + \alpha(2\alpha-1) W_{qa} \\
2(1-\alpha) + \alpha W_{qa} 
\end{matrix} \right\} .
\end{equation}

The first thing we notice is that when $\max W$ in \eqref{eq.maxWobs} is less than 1, it is always given by the first line. This is the relevant case for certifying randomness in practice. Solving for the qubit violation, given an observed violation less than unity we have the bound
\begin{equation}
\label{eq.Wqabound}
W_{qa} \geq \frac{1}{\alpha(2\alpha-1)} [ W - 4\alpha(1-\alpha) ] .
\end{equation}
Second, we note that for $\alpha > 1/2$ the maximum \eqref{eq.maxWobs} is always larger than 1. This means that to be able to certify randomness in practice, we need a minimal fraction of events satisfying the qubit assumption of
\begin{equation}
\label{eq.alphamin}
\alpha > \frac{1}{2} .
\end{equation}
Third, for a given value of $\alpha$ there is a minimal observed violation below which the bound \eqref{eq.Wqabound} becomes trivial and no randomness can be certified. We must have
\begin{equation}
\label{eq.Wobsmin}
W > 4\alpha(1-\alpha) .
\end{equation}

\subsection{Estimating the qubit fraction}

For an implementation with a particular source, we need an estimate or a lower bound on the fraction of qubit events $\alpha$. Source and detector inefficiency, and transmission losses lead to inconclusive events, and our estimate of $\alpha$ should be consistent with how these events are dealt with.

In the scenario of non-malicious, error-prone devices considered here, it is rather natural to discard inconclusive events (e.g. assuming fair-sampling) and then compute $W$ from the remaining data. 
To be able to evaluate \eqref{eq.Wqabound} in this case, one needs to estimate $\alpha$ when inconclusive events are discarded. It is also natural to assume that all events with at most one photon emitted obey the qubit assumption.

With these assumptions, let $q$ denote the probability for the source to emit at most one photon and consider an experiment with $N$ events and $M$ conclusive events. Before post-selection, asymptotically the fraction of events that obey the qubit assumption is then $\alpha = q$. For a finite number of events, we can put a conservative estimate, i.e., a lower bound, on the number of events $N_\alpha$ that satisfy the qubit assumption, within a given confidence. In particular, under the assumption that we know $q$, the behaviour of the source is modelled by a family of $N$ Bernoulli trials parameterized by $q$, and thus the estimation problem can be solved by using the Chernoff-Hoeffding tail inequality. More formally, let $\nu>0$ be the failure probability of the estimation process and $t>0$ be the margin parameter, then
\begin{equation}
\label{eq.nuchernoffdef}
P(N_\alpha \leq qN -t ) \leq \exp(-2Nt^2)=\nu,
\end{equation}
which implies that $N_\alpha > qN- t$ is true with probability at least $1-\nu$. Equivalently, the fraction of qubit events without post-selection is $\alpha > q-t/N$ with probability at least $1-\nu$. The margin parameter $t$ can be expressed in terms of $N$ and $\nu$ as $t=\sqrt{1/(2N)\log(1/\nu)}$.

To account for post-selection, we conservatively assume that all multi-photon events are conclusive. Asymptotically, the fraction of non-qubit events will be $(1-q)N/M$, so $\alpha = 1 - (1-q)N/M$. For finite $N$ we have that after post-selection
\begin{equation}
\alpha \geq 1 - (1-q) \frac{N}{M} - \frac{t}{M}
\end{equation}
with probability at least $1-\nu$, with $\nu$ and $t$ given by \eqref{eq.nuchernoffdef}.

\section{Security Analysis}
\label{app.secanal}
In this section, we show that with the observed experimental statistics, it is possible to provide a bound on the number of random bits that can be extracted from the raw data set, $Z$, which takes values from a set of all binary strings, $\mathcal{Z}$ of length $m$. Our approach essentially uses the (quantum) leftover hash lemma, which states that the amount of private randomness is approximately equal to the min-entropy characterization of the raw data $Z$. More specifically, it says that the number of extractable random bits (that is independent of variables $X,Y,L$) is roughly given by $\Hmin(Z|XYL)$. Here, we recall that variables $X$ and $Y$ are the inputs of Alice and Bob, respectively, and $L$ is the classical register capturing all information about the local variables $\lambda$ and $\mu$. The min-entropy of $Z$ given $XYL$ has a clear operational meaning when casted in terms of the guessing probability, i.e., $\Hmin(Z|XYL)=-m\log_2p_{\t{guess}}$: it measures the probability of correctly guessing $Z$ when given access to classical side-information $XYL$. 

On a more concrete level, the leftover hash lemma employs a family of \textit{universal hash functions} to convert $Z$ into an output string $S$ (of size $\ell$) that is close to a uniform string conditioned on side-information $XYL$.~In particular, we say that the output string $S$ is $\Delta$-\textit{close} to uniform conditioned on $XYL$, if 
\begin{equation}
\frac{1}{2}\sum_{s,x,y,l}|P_{SXYL}-U_SP_{XYL}| \leq \Delta,
\end{equation} where $U_S$ is the uniform distribution of $S$. The quality of the output string is directly related to the number of extractable random bits, i.e., 
\begin{equation}
\ell = \left\lfloor\Hmin(Z|XYL)  - 2\log_2 \frac{1}{2\Delta} \right\rfloor.
\end{equation}
Therefore, to bound $\ell$, we only need to fix a security level $\epsilon_\t{sec}\geq \Delta$ and find a lower bound on the min-entropy term. Using the definition of conditional min-entropy and the assumption that $Z$ is generated from an iid process, we have
\begin{equation}
\ell = \left\lfloor m-m\log_2 \left( 1+  \sqrt{ \frac{1+\sqrt{1-W^2 }}{2}  } \right)  - 2\log_2 \frac{1}{2\Delta} \right\rfloor.
\end{equation} Accordingly, the rate of extraction is $\ell/m$, and it converges to the min-entropy rate when $m \rightarrow \infty$ (therefore $\Delta \rightarrow 0$). At the moment, our bound on $\ell$ is written in terms of the expected value of $W$, which is not directly accessible in the experiment. In order to relate the $W$ to the set of experimental statistics $\mathcal{E}:=\{n^{+}_{x,y}/n_{x,y}\}_{x,y}$, we first use the Chernoff-Hoeffding tail inequality \cite{hoeffding1963}, which provides an upper bound on the probability that the sum of random variables deviates from its expected value. We get 
\begin{equation}
p(1|x,y)-t(\epsilon_\t{pe},n_{x,y})\overset{\epsilon_\t{pe}}{\leq}\frac{n^+_{x,y}}{n_{x,y}} \overset{\epsilon_\t{pe}}{\leq} p(1|xy)+t(\epsilon_\t{pe},n_{x,y}),
\end{equation} where $t(\epsilon_\t{pe},n_{x,y}):=\sqrt{\log(1/\epsilon_\t{pe})/(2n_{x,y})}$. Here, relations with oversetting $\epsilon_\t{pe}$ means that the relations are probabilistically true, i.e., the relations hold except with probability $\epsilon_\t{pe}$. For our purposes later, we denote $p^\pm_{x,y}:=p(1|x,y)\pm t(\epsilon_\t{pe},n_{xy})$. In the following, we introduce an estimate of the expected $W$, i.e., 
\begin{equation}
W \overset{\epsilon'}{\geq}W_\t{min}:=\min_{q_{x,y}\in(p^-_{x,y},p^+_{x,y})}\left| W(\{q_{x,y}\})\right|,
\end{equation}
where $\epsilon'=8\epsilon_{\t{pe}}$ and
\begin{equation}
W(\{q_{x,y}\}):=\det\left[
\begin{matrix}
q_{0,0}-q_{1,0} && q_{2,0}-q_{3,0} \\
q_{0,1}-q_{1,1}&& q_{2,1}-q_{3,1}
\end{matrix}\right] .
\end{equation} Next, we need to bound the maximum fraction of non qubit events, $1-\alpha$. Following the discussion in \secref{app.imperf}, with post-selection we expect $\alpha$ to be $1-\frac{p_2}{p_1+p_2}$ ($p_2$ and $p_1$ are the probabilities of the SPDC to emit, respectively, a double pair or a single-photon pair). In the scenario where $N$ preparations are made, by using the Chernoff-Hoedffing tail inequality, we have that
\begin{equation}
\alpha \overset{\epsilon''}{\geq} \hat{\alpha}:=1-\left[\frac{p_2}{p_1+p_2}+t(\epsilon'',N)\right].
\end{equation} Plugging this into Eq. (C5), we get
\begin{equation}
W_\t{qa} \overset{\epsilon'+\epsilon''}{\geq} \frac{W_\t{min}-4\hat{\alpha}(1-\hat{\alpha})}{\hat{\alpha}(2\hat{\alpha}-1)}.
\end{equation} Therefore, the effective violation is 
\begin{equation}
\hat{W}_{\t{eff}}:=  \frac{W_\t{min}-4\hat{\alpha}(1-\hat{\alpha})}{2\hat{\alpha}-1}.
\end{equation} Note that the effective violation is obtained by fixing the violation due to non qubit contribution to be zero. In other words, the effective violation measures the amount of randomness in $Z$. That is, we have
\[
\ell = \left\lfloor m-m\log_2 \left( 1+  \sqrt{ \frac{1+\sqrt{1-\hat{W}_{\t{eff}}^2 }}{2}  } \right)  - 2\log_2 \frac{1}{2\Delta} \right\rfloor.
\]
Finally, by choosing $\Delta=\epsilon$ and fixing $\epsilon_{\t{pe}}=\epsilon''=\epsilon$, the output string $S$ is $10\epsilon$-\emph{close} to uniform conditioned on $XYL$. In the actual implementation we chose $\epsilon=10^{-3}$.

\begin{figure}[t]
\includegraphics[width=\linewidth]{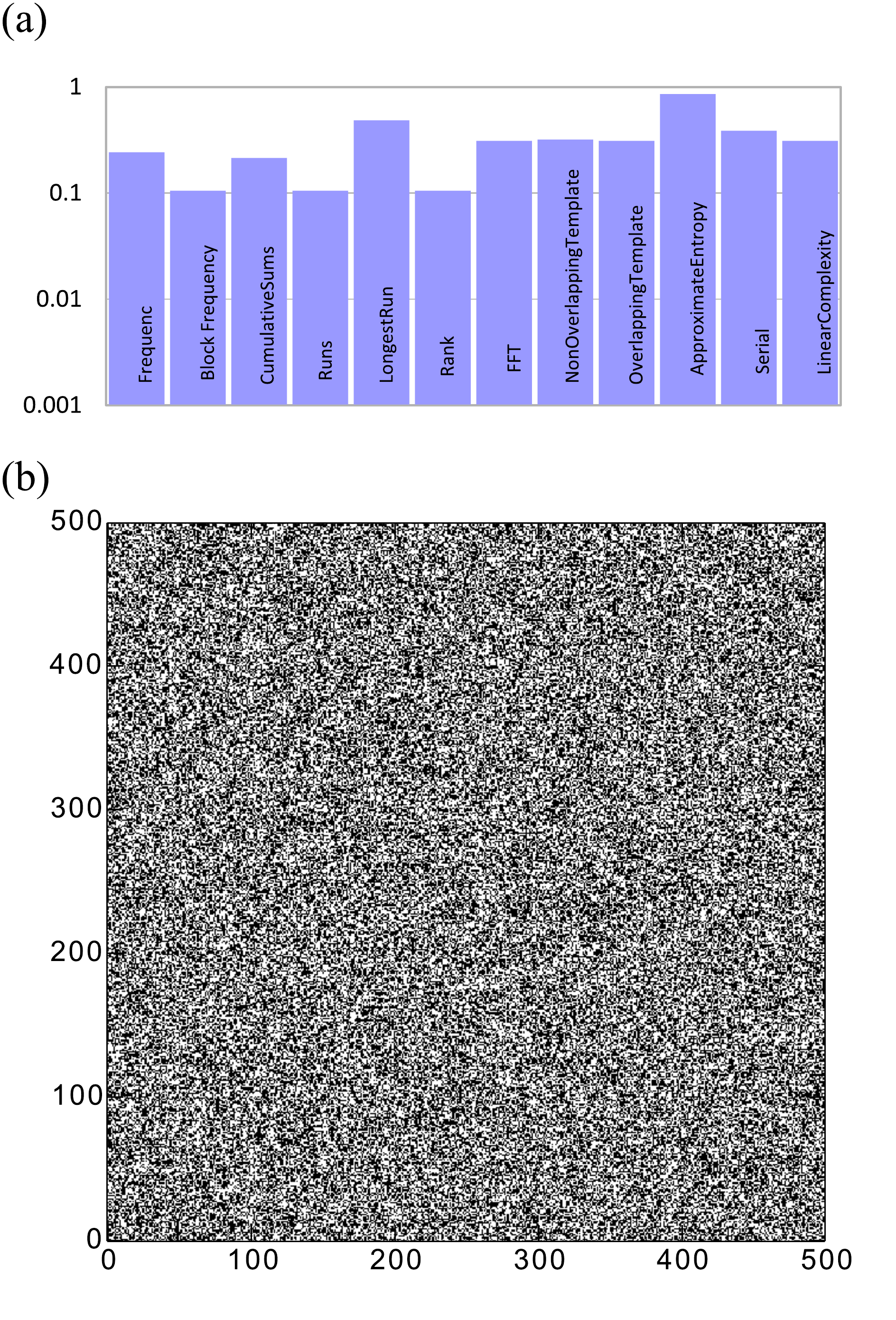}
\caption{(a) NIST tests of the data at the output of the extractor.  (b) Binary image (500$\times$500) of the extracted random bits.}
\label{fig.img}
\end{figure}

\section{Output data analysis}

We performed tests for assessing the quality of the generated randomness, looking for patterns and correlations in the output data. We performed standard statistical test, as defined by NIST. For each test, the p-value is the result of the Kolmogorov-Smirnov test, and must satisfy $0.01 \leq p \leq 0.99$ to be considered successful. Although not all tests could be performed due to the small size of the sample, all performed tests were successful (see \figurename{~\ref{fig.img}-(a)}). A more visual approach to detecting patterns is illustrated in \figurename{~\ref{fig.img}-(b)}, where we display 250000 bits in a 500$\times$500 matrix as a black-and-white image. Any repeated pattern or regular structure in the image would indicate correlations among the bits. No pattern appears.

\section{Example of raw data}

Here, for completeness, we present an extract of the raw data from our experiment, see Tab.~\ref{tab.rawdata}. The data corresponds to one minute of integration, under good alignment conditions. We give the detector counts observed for each detector ($D_1$ and $D_2$), for each measurement setting $y$ and preparation setting $x$. As mentioned in the main text, the preparations $x=\{ 0,1,2,3\}$ correspond respectively to the diagonal (D), anti-diagonal (A), circular right (R) and circular left (L) polarization states. The measurements $y = \{0,1\}$ correspond respectively to the \{D,A\} basis and the \{R,L\} basis. In other words, we use the preparations and measurements of the BB84 protocol.

Based on the raw data, we evaluate the asymptotic probability distribution $p(b|x,y)$ using the method presented in Section IV, and then evaluate the witness value $W$. While perfect BB84 preparations and measurements would give $W=1$ in the asymptotic limit, the observed value is reduced. This is partly due to alignment errors, but especially to finite-size effects. To illustrate, we compute the $W$ value corresponding to the data in Tab.~\ref{tab.rawdata} with and without accounting for finite-size effects. We find $W=0.92$ and $W=0.79$ respectively. These values correspond to visibilities of $V = \sqrt{W} \approx 0.96$ and $V \approx 0.89$ respectively, with respect to the ideal BB84 preparations and measurements mixed with white noise. Note that $W=0.79$ is not far from the average $W=0.76$ observed under good conditions (see the main text).

\begin{table}
\begin{tabular}{cc|c|c|c|c|}
\cline{3-6}
& &  \multicolumn{4}{c|}{Measurement} \\
\cline{3-6}
& &  \multicolumn{2}{c|}{D/A} &  \multicolumn{2}{c|}{R/L} \\
\cline{3-6}
& & $D_1$ & $D_2$ & $D_1$ & $D_2$ \\
\hline
\multicolumn{1}{ |c  }{ \multirow{4}{*}{\rotatebox[origin=c]{90}{Preparation\hspace{0.25cm}}} } & \multicolumn{1}{ |c| }{\parbox[c][2em]{0.2cm}{D}} & 5903 & 97 & 3515 & 2485 \\[1ex]
\cline{2-6}
\multicolumn{1}{ |c  }{} & \multicolumn{1}{ |c| }{\parbox[c][2em]{0.2cm}{A}} & 172 & 5828 & 2950 & 3050 \\[1ex]
\cline{2-6}
\multicolumn{1}{ |c  }{} & \multicolumn{1}{ |c| }{\parbox[c][2em]{0.2cm}{R}} & 2825 & 3175 & 5914 & 86 \\[1ex]
\cline{2-6}
\multicolumn{1}{ |c  }{} & \multicolumn{1}{ |c| }{\parbox[c][2em]{0.2cm}{L}} & 3565 & 2435 & 199 & 5801 \\[1ex]
\hline
\end{tabular}
\caption{Sample of raw data taken during one minute under good alignment conditions.}\label{tab.rawdata}
\end{table}


\begin{thebibliography}{36}%
\makeatletter
\providecommand \@ifxundefined [1]{%
 \@ifx{#1\undefined}
}%
\providecommand \@ifnum [1]{%
 \ifnum #1\expandafter \@firstoftwo
 \else \expandafter \@secondoftwo
 \fi
}%
\providecommand \@ifx [1]{%
 \ifx #1\expandafter \@firstoftwo
 \else \expandafter \@secondoftwo
 \fi
}%
\providecommand \natexlab [1]{#1}%
\providecommand \enquote  [1]{``#1''}%
\providecommand \bibnamefont  [1]{#1}%
\providecommand \bibfnamefont [1]{#1}%
\providecommand \citenamefont [1]{#1}%
\providecommand \href@noop [0]{\@secondoftwo}%
\providecommand \href [0]{\begingroup \@sanitize@url \@href}%
\providecommand \@href[1]{\@@startlink{#1}\@@href}%
\providecommand \@@href[1]{\endgroup#1\@@endlink}%
\providecommand \@sanitize@url [0]{\catcode `\\12\catcode `\$12\catcode
  `\&12\catcode `\#12\catcode `\^12\catcode `\_12\catcode `\%12\relax}%
\providecommand \@@startlink[1]{}%
\providecommand \@@endlink[0]{}%
\providecommand \url  [0]{\begingroup\@sanitize@url \@url }%
\providecommand \@url [1]{\endgroup\@href {#1}{\urlprefix }}%
\providecommand \urlprefix  [0]{URL }%
\providecommand \Eprint [0]{\href }%
\providecommand \doibase [0]{http://dx.doi.org/}%
\providecommand \selectlanguage [0]{\@gobble}%
\providecommand \bibinfo  [0]{\@secondoftwo}%
\providecommand \bibfield  [0]{\@secondoftwo}%
\providecommand \translation [1]{[#1]}%
\providecommand \BibitemOpen [0]{}%
\providecommand \bibitemStop [0]{}%
\providecommand \bibitemNoStop [0]{.\EOS\space}%
\providecommand \EOS [0]{\spacefactor3000\relax}%
\providecommand \BibitemShut  [1]{\csname bibitem#1\endcsname}%
\let\auto@bib@innerbib\@empty
\bibitem [{\citenamefont {Rarity}\ \emph {et~al.}(1994)\citenamefont {Rarity},
  \citenamefont {Owens},\ and\ \citenamefont {Tapster}}]{rarity1994}%
  \BibitemOpen
  \bibfield  {author} {\bibinfo {author} {\bibfnamefont {J.}~\bibnamefont
  {Rarity}}, \bibinfo {author} {\bibfnamefont {P.}~\bibnamefont {Owens}}, \
  and\ \bibinfo {author} {\bibfnamefont {P.}~\bibnamefont {Tapster}},\ }\href
  {\doibase 10.1080/09500349414552281} {\bibfield  {journal} {\bibinfo
  {journal} {Journal of Modern Optics}\ }\textbf {\bibinfo {volume} {41}},\
  \bibinfo {pages} {2435} (\bibinfo {year} {1994})}\BibitemShut {NoStop}%
\bibitem [{\citenamefont {Stefanov}\ \emph {et~al.}(2000)\citenamefont
  {Stefanov}, \citenamefont {Gisin}, \citenamefont {Guinnard}, \citenamefont
  {Guinnard},\ and\ \citenamefont {Zbinden}}]{stefanov2000}%
  \BibitemOpen
  \bibfield  {author} {\bibinfo {author} {\bibfnamefont {A.}~\bibnamefont
  {Stefanov}}, \bibinfo {author} {\bibfnamefont {N.}~\bibnamefont {Gisin}},
  \bibinfo {author} {\bibfnamefont {O.}~\bibnamefont {Guinnard}}, \bibinfo
  {author} {\bibfnamefont {L.}~\bibnamefont {Guinnard}}, \ and\ \bibinfo
  {author} {\bibfnamefont {H.}~\bibnamefont {Zbinden}},\ }\href
  {http://dx.doi.org/10.1080/09500340008233380} {\bibfield  {journal} {\bibinfo
   {journal} {Journal of Modern Optics}\ }\textbf {\bibinfo {volume} {47}},\
  \bibinfo {pages} {595} (\bibinfo {year} {2000})}\BibitemShut {NoStop}%
\bibitem [{\citenamefont {Jennewein}\ \emph {et~al.}(2000)\citenamefont
  {Jennewein}, \citenamefont {Achleitner}, \citenamefont {Weihs}, \citenamefont
  {Weinfurter},\ and\ \citenamefont {Zeilinger}}]{jennewein2000}%
  \BibitemOpen
  \bibfield  {author} {\bibinfo {author} {\bibfnamefont {T.}~\bibnamefont
  {Jennewein}}, \bibinfo {author} {\bibfnamefont {U.}~\bibnamefont
  {Achleitner}}, \bibinfo {author} {\bibfnamefont {G.}~\bibnamefont {Weihs}},
  \bibinfo {author} {\bibfnamefont {H.}~\bibnamefont {Weinfurter}}, \ and\
  \bibinfo {author} {\bibfnamefont {A.}~\bibnamefont {Zeilinger}},\ }\href
  {\doibase http://dx.doi.org/10.1063/1.1150518} {\bibfield  {journal}
  {\bibinfo  {journal} {Review of Scientific Instruments}\ }\textbf {\bibinfo
  {volume} {71}},\ \bibinfo {pages} {1675} (\bibinfo {year}
  {2000})}\BibitemShut {NoStop}%
\bibitem [{\citenamefont {Dynes}\ \emph {et~al.}(2008)\citenamefont {Dynes},
  \citenamefont {Yuan}, \citenamefont {Sharpe},\ and\ \citenamefont
  {Shields}}]{dynes2008}%
  \BibitemOpen
  \bibfield  {author} {\bibinfo {author} {\bibfnamefont {J.~F.}\ \bibnamefont
  {Dynes}}, \bibinfo {author} {\bibfnamefont {Z.~L.}\ \bibnamefont {Yuan}},
  \bibinfo {author} {\bibfnamefont {A.~W.}\ \bibnamefont {Sharpe}}, \ and\
  \bibinfo {author} {\bibfnamefont {A.~J.}\ \bibnamefont {Shields}},\ }\href
  {\doibase http://dx.doi.org/10.1063/1.2961000} {\bibfield  {journal}
  {\bibinfo  {journal} {Applied Physics Letters}\ }\textbf {\bibinfo {volume}
  {93}},\ \bibinfo {eid} {031109} (\bibinfo {year} {2008})}\BibitemShut
  {NoStop}%
\bibitem [{\citenamefont {Wahl}\ \emph {et~al.}(2011)\citenamefont {Wahl},
  \citenamefont {Leifgen}, \citenamefont {Berlin}, \citenamefont {Röhlicke},
  \citenamefont {Rahn},\ and\ \citenamefont {Benson}}]{wahl2011}%
  \BibitemOpen
  \bibfield  {author} {\bibinfo {author} {\bibfnamefont {M.}~\bibnamefont
  {Wahl}}, \bibinfo {author} {\bibfnamefont {M.}~\bibnamefont {Leifgen}},
  \bibinfo {author} {\bibfnamefont {M.}~\bibnamefont {Berlin}}, \bibinfo
  {author} {\bibfnamefont {T.}~\bibnamefont {Röhlicke}}, \bibinfo {author}
  {\bibfnamefont {H.-J.}\ \bibnamefont {Rahn}}, \ and\ \bibinfo {author}
  {\bibfnamefont {O.}~\bibnamefont {Benson}},\ }\href {\doibase
  http://dx.doi.org/10.1063/1.3578456} {\bibfield  {journal} {\bibinfo
  {journal} {Applied Physics Letters}\ }\textbf {\bibinfo {volume} {98}},\
  \bibinfo {eid} {171105} (\bibinfo {year} {2011})}\BibitemShut {NoStop}%
\bibitem [{\citenamefont {Nie}\ \emph {et~al.}(2014)\citenamefont {Nie},
  \citenamefont {Zhang}, \citenamefont {Zhang}, \citenamefont {Wang},
  \citenamefont {Ma}, \citenamefont {Zhang},\ and\ \citenamefont
  {Pan}}]{nie2014}%
  \BibitemOpen
  \bibfield  {author} {\bibinfo {author} {\bibfnamefont {Y.-Q.}\ \bibnamefont
  {Nie}}, \bibinfo {author} {\bibfnamefont {H.-F.}\ \bibnamefont {Zhang}},
  \bibinfo {author} {\bibfnamefont {Z.}~\bibnamefont {Zhang}}, \bibinfo
  {author} {\bibfnamefont {J.}~\bibnamefont {Wang}}, \bibinfo {author}
  {\bibfnamefont {X.}~\bibnamefont {Ma}}, \bibinfo {author} {\bibfnamefont
  {J.}~\bibnamefont {Zhang}}, \ and\ \bibinfo {author} {\bibfnamefont {J.-W.}\
  \bibnamefont {Pan}},\ }\href {\doibase http://dx.doi.org/10.1063/1.4863224}
  {\bibfield  {journal} {\bibinfo  {journal} {Applied Physics Letters}\
  }\textbf {\bibinfo {volume} {104}},\ \bibinfo {eid} {051110} (\bibinfo {year}
  {2014})}\BibitemShut {NoStop}%
\bibitem [{\citenamefont {Stip\v{c}evi\'c}\ and\ \citenamefont
  {Rogina}(2007)}]{stipcevic2007}%
  \BibitemOpen
  \bibfield  {author} {\bibinfo {author} {\bibfnamefont {M.}~\bibnamefont
  {Stip\v{c}evi\'c}}\ and\ \bibinfo {author} {\bibfnamefont {B.~M.}\
  \bibnamefont {Rogina}},\ }\href {\doibase
  http://dx.doi.org/10.1063/1.2720728} {\bibfield  {journal} {\bibinfo
  {journal} {Review of Scientific Instruments}\ }\textbf {\bibinfo {volume}
  {78}},\ \bibinfo {eid} {045104} (\bibinfo {year} {2007})}\BibitemShut
  {NoStop}%
\bibitem [{\citenamefont {Qi}\ \emph {et~al.}(2010)\citenamefont {Qi},
  \citenamefont {Chi}, \citenamefont {Lo},\ and\ \citenamefont
  {Qian}}]{qi2010}%
  \BibitemOpen
  \bibfield  {author} {\bibinfo {author} {\bibfnamefont {B.}~\bibnamefont
  {Qi}}, \bibinfo {author} {\bibfnamefont {Y.-M.}\ \bibnamefont {Chi}},
  \bibinfo {author} {\bibfnamefont {H.-K.}\ \bibnamefont {Lo}}, \ and\ \bibinfo
  {author} {\bibfnamefont {L.}~\bibnamefont {Qian}},\ }\href {\doibase
  10.1364/OL.35.000312} {\bibfield  {journal} {\bibinfo  {journal} {Opt.
  Lett.}\ }\textbf {\bibinfo {volume} {35}},\ \bibinfo {pages} {312} (\bibinfo
  {year} {2010})}\BibitemShut {NoStop}%
\bibitem [{\citenamefont {Uchida}\ \emph {et~al.}(2008)\citenamefont {Uchida},
  \citenamefont {Amano}, \citenamefont {Inoue}, \citenamefont {Hirano},
  \citenamefont {Naito}, \citenamefont {Someya}, \citenamefont {OowadaIsao},
  \citenamefont {Kurashige}, \citenamefont {Shiki}, \citenamefont {Yoshimori},
  \citenamefont {Yoshimura},\ and\ \citenamefont {Davis}}]{uchida2008}%
  \BibitemOpen
  \bibfield  {author} {\bibinfo {author} {\bibfnamefont {A.}~\bibnamefont
  {Uchida}}, \bibinfo {author} {\bibfnamefont {K.}~\bibnamefont {Amano}},
  \bibinfo {author} {\bibfnamefont {M.}~\bibnamefont {Inoue}}, \bibinfo
  {author} {\bibfnamefont {K.}~\bibnamefont {Hirano}}, \bibinfo {author}
  {\bibfnamefont {S.}~\bibnamefont {Naito}}, \bibinfo {author} {\bibfnamefont
  {H.}~\bibnamefont {Someya}}, \bibinfo {author} {\bibnamefont {OowadaIsao}},
  \bibinfo {author} {\bibfnamefont {T.}~\bibnamefont {Kurashige}}, \bibinfo
  {author} {\bibfnamefont {M.}~\bibnamefont {Shiki}}, \bibinfo {author}
  {\bibfnamefont {S.}~\bibnamefont {Yoshimori}}, \bibinfo {author}
  {\bibfnamefont {K.}~\bibnamefont {Yoshimura}}, \ and\ \bibinfo {author}
  {\bibfnamefont {P.}~\bibnamefont {Davis}},\ }\href
  {http://dx.doi.org/10.1038/nphoton.2008.227} {\bibfield  {journal} {\bibinfo
  {journal} {Nat Photon}\ }\textbf {\bibinfo {volume} {2}},\ \bibinfo {pages}
  {728} (\bibinfo {year} {2008})}\BibitemShut {NoStop}%
\bibitem [{\citenamefont {Abell\'{a}n}\ \emph {et~al.}(2014)\citenamefont
  {Abell\'{a}n}, \citenamefont {Amaya}, \citenamefont {Jofre}, \citenamefont
  {Curty}, \citenamefont {Ac\'{i}n}, \citenamefont {Capmany}, \citenamefont
  {Pruneri},\ and\ \citenamefont {Mitchell}}]{abellan2014}%
  \BibitemOpen
  \bibfield  {author} {\bibinfo {author} {\bibfnamefont {C.}~\bibnamefont
  {Abell\'{a}n}}, \bibinfo {author} {\bibfnamefont {W.}~\bibnamefont {Amaya}},
  \bibinfo {author} {\bibfnamefont {M.}~\bibnamefont {Jofre}}, \bibinfo
  {author} {\bibfnamefont {M.}~\bibnamefont {Curty}}, \bibinfo {author}
  {\bibfnamefont {A.}~\bibnamefont {Ac\'{i}n}}, \bibinfo {author}
  {\bibfnamefont {J.}~\bibnamefont {Capmany}}, \bibinfo {author} {\bibfnamefont
  {V.}~\bibnamefont {Pruneri}}, \ and\ \bibinfo {author} {\bibfnamefont
  {M.~W.}\ \bibnamefont {Mitchell}},\ }\href {\doibase 10.1364/OE.22.001645}
  {\bibfield  {journal} {\bibinfo  {journal} {Opt. Express}\ }\textbf {\bibinfo
  {volume} {22}},\ \bibinfo {pages} {1645} (\bibinfo {year}
  {2014})}\BibitemShut {NoStop}%
\bibitem [{\citenamefont {Gabriel}\ \emph {et~al.}(2010)\citenamefont
  {Gabriel}, \citenamefont {Wittmann}, \citenamefont {Sych}, \citenamefont
  {Dong}, \citenamefont {Mauerer}, \citenamefont {Andersen}, \citenamefont
  {Marquardt},\ and\ \citenamefont {Leuchs}}]{gabriel2010}%
  \BibitemOpen
  \bibfield  {author} {\bibinfo {author} {\bibfnamefont {C.}~\bibnamefont
  {Gabriel}}, \bibinfo {author} {\bibfnamefont {C.}~\bibnamefont {Wittmann}},
  \bibinfo {author} {\bibfnamefont {D.}~\bibnamefont {Sych}}, \bibinfo {author}
  {\bibfnamefont {R.}~\bibnamefont {Dong}}, \bibinfo {author} {\bibfnamefont
  {W.}~\bibnamefont {Mauerer}}, \bibinfo {author} {\bibfnamefont {U.~L.}\
  \bibnamefont {Andersen}}, \bibinfo {author} {\bibfnamefont {C.}~\bibnamefont
  {Marquardt}}, \ and\ \bibinfo {author} {\bibfnamefont {G.}~\bibnamefont
  {Leuchs}},\ }\href {http://dx.doi.org/10.1038/nphoton.2010.197} {\bibfield
  {journal} {\bibinfo  {journal} {Nat Photon}\ }\textbf {\bibinfo {volume}
  {4}},\ \bibinfo {pages} {711} (\bibinfo {year} {2010})}\BibitemShut {NoStop}%
\bibitem [{\citenamefont {Symul}\ \emph {et~al.}(2011)\citenamefont {Symul},
  \citenamefont {Assad},\ and\ \citenamefont {Lam}}]{symul2011}%
  \BibitemOpen
  \bibfield  {author} {\bibinfo {author} {\bibfnamefont {T.}~\bibnamefont
  {Symul}}, \bibinfo {author} {\bibfnamefont {S.~M.}\ \bibnamefont {Assad}}, \
  and\ \bibinfo {author} {\bibfnamefont {P.~K.}\ \bibnamefont {Lam}},\ }\href
  {\doibase http://dx.doi.org/10.1063/1.3597793} {\bibfield  {journal}
  {\bibinfo  {journal} {Applied Physics Letters}\ }\textbf {\bibinfo {volume}
  {98}},\ \bibinfo {eid} {231103} (\bibinfo {year} {2011})}\BibitemShut
  {NoStop}%
\bibitem [{\citenamefont {Sanguinetti}\ \emph {et~al.}(2014)\citenamefont
  {Sanguinetti}, \citenamefont {Martin}, \citenamefont {Zbinden},\ and\
  \citenamefont {Gisin}}]{sanguinetti2014}%
  \BibitemOpen
  \bibfield  {author} {\bibinfo {author} {\bibfnamefont {B.}~\bibnamefont
  {Sanguinetti}}, \bibinfo {author} {\bibfnamefont {A.}~\bibnamefont {Martin}},
  \bibinfo {author} {\bibfnamefont {H.}~\bibnamefont {Zbinden}}, \ and\
  \bibinfo {author} {\bibfnamefont {N.}~\bibnamefont {Gisin}},\ }\href
  {\doibase 10.1103/PhysRevX.4.031056} {\bibfield  {journal} {\bibinfo
  {journal} {Phys. Rev. X}\ }\textbf {\bibinfo {volume} {4}},\ \bibinfo {pages}
  {031056} (\bibinfo {year} {2014})}\BibitemShut {NoStop}%
\bibitem [{\citenamefont {Nisan}\ and\ \citenamefont
  {Ta-Shma}(1999)}]{nisan1999}%
  \BibitemOpen
  \bibfield  {author} {\bibinfo {author} {\bibfnamefont {N.}~\bibnamefont
  {Nisan}}\ and\ \bibinfo {author} {\bibfnamefont {A.}~\bibnamefont
  {Ta-Shma}},\ }\href {\doibase http://dx.doi.org/10.1006/jcss.1997.1546}
  {\bibfield  {journal} {\bibinfo  {journal} {Journal of Computer and System
  Sciences}\ }\textbf {\bibinfo {volume} {58}},\ \bibinfo {pages} {148 }
  (\bibinfo {year} {1999})}\BibitemShut {NoStop}%
\bibitem [{\citenamefont {Dodis}\ \emph {et~al.}(2013)\citenamefont {Dodis},
  \citenamefont {Pointcheval}, \citenamefont {Ruhault}, \citenamefont
  {Vergniaud},\ and\ \citenamefont {Wichs}}]{dodis2013}%
  \BibitemOpen
  \bibfield  {author} {\bibinfo {author} {\bibfnamefont {Y.}~\bibnamefont
  {Dodis}}, \bibinfo {author} {\bibfnamefont {D.}~\bibnamefont {Pointcheval}},
  \bibinfo {author} {\bibfnamefont {S.}~\bibnamefont {Ruhault}}, \bibinfo
  {author} {\bibfnamefont {D.}~\bibnamefont {Vergniaud}}, \ and\ \bibinfo
  {author} {\bibfnamefont {D.}~\bibnamefont {Wichs}},\ }\href {\doibase
  10.1145/2508859.2516653} {\bibfield  {journal} {\bibinfo  {journal}
  {Proceedings of the 2013 ACM SIGSAC Conference on Computer \& Communications
  Security}\ }\bibinfo {series} {CCS '13},\ \bibinfo {pages} {647} (\bibinfo
  {year} {2013})}\BibitemShut {NoStop}%
\bibitem [{\citenamefont {Frauchiger}\ \emph {et~al.}(2013)\citenamefont
  {Frauchiger}, \citenamefont {Renner},\ and\ \citenamefont
  {Troyer}}]{frauchiger2013}%
  \BibitemOpen
  \bibfield  {author} {\bibinfo {author} {\bibfnamefont {D.}~\bibnamefont
  {Frauchiger}}, \bibinfo {author} {\bibfnamefont {R.}~\bibnamefont {Renner}},
  \ and\ \bibinfo {author} {\bibfnamefont {M.}~\bibnamefont {Troyer}},\
  }\href@noop {} {\bibfield  {journal} {\bibinfo  {journal} {arXiv e-print}\ }
  (\bibinfo {year} {2013})},\ \Eprint {http://arxiv.org/abs/1311.4547}
  {arXiv:1311.4547 [quant-ph]} \BibitemShut {NoStop}%
\bibitem [{\citenamefont {Ma}\ \emph {et~al.}(2013)\citenamefont {Ma},
  \citenamefont {Xu}, \citenamefont {Xu}, \citenamefont {Tan}, \citenamefont
  {Qi},\ and\ \citenamefont {Lo}}]{ma2013}%
  \BibitemOpen
  \bibfield  {author} {\bibinfo {author} {\bibfnamefont {X.}~\bibnamefont
  {Ma}}, \bibinfo {author} {\bibfnamefont {F.}~\bibnamefont {Xu}}, \bibinfo
  {author} {\bibfnamefont {H.}~\bibnamefont {Xu}}, \bibinfo {author}
  {\bibfnamefont {X.}~\bibnamefont {Tan}}, \bibinfo {author} {\bibfnamefont
  {B.}~\bibnamefont {Qi}}, \ and\ \bibinfo {author} {\bibfnamefont {H.-K.}\
  \bibnamefont {Lo}},\ }\href {\doibase 10.1103/PhysRevA.87.062327} {\bibfield
  {journal} {\bibinfo  {journal} {Phys. Rev. A}\ }\textbf {\bibinfo {volume}
  {87}},\ \bibinfo {pages} {062327} (\bibinfo {year} {2013})}\BibitemShut
  {NoStop}%
\bibitem [{\citenamefont {Colbeck}()}]{colbeckPhD}%
  \BibitemOpen
  \bibfield  {author} {\bibinfo {author} {\bibfnamefont {R.}~\bibnamefont
  {Colbeck}},\ }\href {http://arxiv.org/abs/0911.3814} {\enquote {\bibinfo
  {title} {Ph.d. thesis, arxiv:0911.3814},}\ }\BibitemShut {NoStop}%
\bibitem [{\citenamefont {Pironio}\ \emph {et~al.}(2010)\citenamefont
  {Pironio}, \citenamefont {Ac\'in}, \citenamefont {Massar}, \citenamefont
  {de~la Giroday}, \citenamefont {Matsukevich}, \citenamefont {Maunz},
  \citenamefont {Olmschenk}, \citenamefont {Hayes}, \citenamefont {Luo},
  \citenamefont {Manning},\ and\ \citenamefont {Monroe}}]{pironio2010}%
  \BibitemOpen
  \bibfield  {author} {\bibinfo {author} {\bibfnamefont {S.}~\bibnamefont
  {Pironio}}, \bibinfo {author} {\bibfnamefont {A.}~\bibnamefont {Ac\'in}},
  \bibinfo {author} {\bibfnamefont {S.}~\bibnamefont {Massar}}, \bibinfo
  {author} {\bibfnamefont {A.~B.}\ \bibnamefont {de~la Giroday}}, \bibinfo
  {author} {\bibfnamefont {D.~N.}\ \bibnamefont {Matsukevich}}, \bibinfo
  {author} {\bibfnamefont {P.}~\bibnamefont {Maunz}}, \bibinfo {author}
  {\bibfnamefont {S.}~\bibnamefont {Olmschenk}}, \bibinfo {author}
  {\bibfnamefont {D.}~\bibnamefont {Hayes}}, \bibinfo {author} {\bibfnamefont
  {L.}~\bibnamefont {Luo}}, \bibinfo {author} {\bibfnamefont {T.~A.}\
  \bibnamefont {Manning}}, \ and\ \bibinfo {author} {\bibfnamefont
  {C.}~\bibnamefont {Monroe}},\ }\href {http://dx.doi.org/10.1038/nature09008}
  {\bibfield  {journal} {\bibinfo  {journal} {Nature}\ }\textbf {\bibinfo
  {volume} {464}},\ \bibinfo {pages} {1021} (\bibinfo {year}
  {2010})}\BibitemShut {NoStop}%
\bibitem [{\citenamefont {Christensen}\ \emph {et~al.}(2013)\citenamefont
  {Christensen}, \citenamefont {McCusker}, \citenamefont {Altepeter},
  \citenamefont {Calkins}, \citenamefont {Gerrits}, \citenamefont {Lita},
  \citenamefont {Miller}, \citenamefont {Shalm}, \citenamefont {Zhang},
  \citenamefont {Nam}, \citenamefont {Brunner}, \citenamefont {Lim},
  \citenamefont {Gisin},\ and\ \citenamefont {Kwiat}}]{christensen2013}%
  \BibitemOpen
  \bibfield  {author} {\bibinfo {author} {\bibfnamefont {B.~G.}\ \bibnamefont
  {Christensen}}, \bibinfo {author} {\bibfnamefont {K.~T.}\ \bibnamefont
  {McCusker}}, \bibinfo {author} {\bibfnamefont {J.~B.}\ \bibnamefont
  {Altepeter}}, \bibinfo {author} {\bibfnamefont {B.}~\bibnamefont {Calkins}},
  \bibinfo {author} {\bibfnamefont {T.}~\bibnamefont {Gerrits}}, \bibinfo
  {author} {\bibfnamefont {A.~E.}\ \bibnamefont {Lita}}, \bibinfo {author}
  {\bibfnamefont {A.}~\bibnamefont {Miller}}, \bibinfo {author} {\bibfnamefont
  {L.~K.}\ \bibnamefont {Shalm}}, \bibinfo {author} {\bibfnamefont
  {Y.}~\bibnamefont {Zhang}}, \bibinfo {author} {\bibfnamefont {S.~W.}\
  \bibnamefont {Nam}}, \bibinfo {author} {\bibfnamefont {N.}~\bibnamefont
  {Brunner}}, \bibinfo {author} {\bibfnamefont {C.~C.~W.}\ \bibnamefont {Lim}},
  \bibinfo {author} {\bibfnamefont {N.}~\bibnamefont {Gisin}}, \ and\ \bibinfo
  {author} {\bibfnamefont {P.~G.}\ \bibnamefont {Kwiat}},\ }\href {\doibase
  10.1103/PhysRevLett.111.130406} {\bibfield  {journal} {\bibinfo  {journal}
  {Phys. Rev. Lett.}\ }\textbf {\bibinfo {volume} {111}},\ \bibinfo {pages}
  {130406} (\bibinfo {year} {2013})}\BibitemShut {NoStop}%
\bibitem [{\citenamefont {Li}\ \emph {et~al.}(2011)\citenamefont {Li},
  \citenamefont {Yin}, \citenamefont {Wu}, \citenamefont {Zou}, \citenamefont
  {Wang}, \citenamefont {Chen}, \citenamefont {Guo},\ and\ \citenamefont
  {Han}}]{li2011}%
  \BibitemOpen
  \bibfield  {author} {\bibinfo {author} {\bibfnamefont {H.-W.}\ \bibnamefont
  {Li}}, \bibinfo {author} {\bibfnamefont {Z.-Q.}\ \bibnamefont {Yin}},
  \bibinfo {author} {\bibfnamefont {Y.-C.}\ \bibnamefont {Wu}}, \bibinfo
  {author} {\bibfnamefont {X.-B.}\ \bibnamefont {Zou}}, \bibinfo {author}
  {\bibfnamefont {S.}~\bibnamefont {Wang}}, \bibinfo {author} {\bibfnamefont
  {W.}~\bibnamefont {Chen}}, \bibinfo {author} {\bibfnamefont {G.-C.}\
  \bibnamefont {Guo}}, \ and\ \bibinfo {author} {\bibfnamefont {Z.-F.}\
  \bibnamefont {Han}},\ }\href {\doibase 10.1103/PhysRevA.84.034301} {\bibfield
   {journal} {\bibinfo  {journal} {Phys. Rev. A}\ }\textbf {\bibinfo {volume}
  {84}},\ \bibinfo {pages} {034301} (\bibinfo {year} {2011})}\BibitemShut
  {NoStop}%
\bibitem [{\citenamefont {Li}\ \emph {et~al.}(2012)\citenamefont {Li},
  \citenamefont {Paw\l{}owski}, \citenamefont {Yin}, \citenamefont {Guo},\ and\
  \citenamefont {Han}}]{li2012}%
  \BibitemOpen
  \bibfield  {author} {\bibinfo {author} {\bibfnamefont {H.-W.}\ \bibnamefont
  {Li}}, \bibinfo {author} {\bibfnamefont {M.}~\bibnamefont {Paw\l{}owski}},
  \bibinfo {author} {\bibfnamefont {Z.-Q.}\ \bibnamefont {Yin}}, \bibinfo
  {author} {\bibfnamefont {G.-C.}\ \bibnamefont {Guo}}, \ and\ \bibinfo
  {author} {\bibfnamefont {Z.-F.}\ \bibnamefont {Han}},\ }\href {\doibase
  10.1103/PhysRevA.85.052308} {\bibfield  {journal} {\bibinfo  {journal} {Phys.
  Rev. A}\ }\textbf {\bibinfo {volume} {85}},\ \bibinfo {pages} {052308}
  (\bibinfo {year} {2012})}\BibitemShut {NoStop}%
\bibitem [{\citenamefont {Dall'Arno}\ \emph {et~al.}(2012)\citenamefont
  {Dall'Arno}, \citenamefont {Passaro}, \citenamefont {Gallego}, \citenamefont
  {Pawlowski},\ and\ \citenamefont {Acin}}]{dallArno2012}%
  \BibitemOpen
  \bibfield  {author} {\bibinfo {author} {\bibfnamefont {M.}~\bibnamefont
  {Dall'Arno}}, \bibinfo {author} {\bibfnamefont {E.}~\bibnamefont {Passaro}},
  \bibinfo {author} {\bibfnamefont {R.}~\bibnamefont {Gallego}}, \bibinfo
  {author} {\bibfnamefont {M.}~\bibnamefont {Pawlowski}}, \ and\ \bibinfo
  {author} {\bibfnamefont {A.}~\bibnamefont {Acin}},\ }\href
  {http://arxiv.org/abs/1210.1272} {\bibfield  {journal} {\bibinfo  {journal}
  {arXiv e-print}\ } (\bibinfo {year} {2012})},\ \Eprint
  {http://arxiv.org/abs/1210.1272} {arXiv:1210.1272 [quant-ph]} \BibitemShut
  {NoStop}%
\bibitem [{\citenamefont {Vallone}\ \emph {et~al.}(2014)\citenamefont
  {Vallone}, \citenamefont {Marangon}, \citenamefont {Tomasin},\ and\
  \citenamefont {PaoloVilloresi}}]{vallone2014}%
  \BibitemOpen
  \bibfield  {author} {\bibinfo {author} {\bibfnamefont {G.}~\bibnamefont
  {Vallone}}, \bibinfo {author} {\bibfnamefont {D.~G.}\ \bibnamefont
  {Marangon}}, \bibinfo {author} {\bibfnamefont {M.}~\bibnamefont {Tomasin}}, \
  and\ \bibinfo {author} {\bibnamefont {PaoloVilloresi}},\ }\href@noop {}
  {\bibfield  {journal} {\bibinfo  {journal} {arXiv e-print}\ } (\bibinfo
  {year} {2014})},\ \Eprint {http://arxiv.org/abs/1401.7917} {arXiv:1401.7917
  [quant-ph]} \BibitemShut {NoStop}%
\bibitem [{\citenamefont {Bowles}\ \emph {et~al.}(2014)\citenamefont {Bowles},
  \citenamefont {Quintino},\ and\ \citenamefont {Brunner}}]{bowles2013}%
  \BibitemOpen
  \bibfield  {author} {\bibinfo {author} {\bibfnamefont {J.}~\bibnamefont
  {Bowles}}, \bibinfo {author} {\bibfnamefont {M.~T.}\ \bibnamefont
  {Quintino}}, \ and\ \bibinfo {author} {\bibfnamefont {N.}~\bibnamefont
  {Brunner}},\ }\href {\doibase 10.1103/PhysRevLett.112.140407} {\bibfield
  {journal} {\bibinfo  {journal} {Phys. Rev. Lett.}\ }\textbf {\bibinfo
  {volume} {112}},\ \bibinfo {pages} {140407} (\bibinfo {year}
  {2014})}\BibitemShut {NoStop}%
\bibitem [{sup()}]{suppinfo}%
  \BibitemOpen
  \href@noop {} {\enquote {\bibinfo {title} {See {Supplementary Material}},}\ }\BibitemShut {NoStop}%
\bibitem [{\citenamefont {Koenig}\ \emph {et~al.}(2009)\citenamefont {Koenig},
  \citenamefont {Renner},\ and\ \citenamefont {Schaffner}}]{koenig2009}%
  \BibitemOpen
  \bibfield  {author} {\bibinfo {author} {\bibfnamefont {R.}~\bibnamefont
  {Koenig}}, \bibinfo {author} {\bibfnamefont {R.}~\bibnamefont {Renner}}, \
  and\ \bibinfo {author} {\bibfnamefont {C.}~\bibnamefont {Schaffner}},\ }\href
  {\doibase 10.1109/TIT.2009.2025545} {\bibfield  {journal} {\bibinfo
  {journal} {Information Theory, IEEE Transactions on}\ }\textbf {\bibinfo
  {volume} {55}},\ \bibinfo {pages} {4337} (\bibinfo {year}
  {2009})}\BibitemShut {NoStop}%
\bibitem [{\citenamefont {Bennett}\ and\ \citenamefont
  {Brassard}(1984)}]{bennett1984}%
  \BibitemOpen
  \bibfield  {author} {\bibinfo {author} {\bibfnamefont {C.~H.}\ \bibnamefont
  {Bennett}}\ and\ \bibinfo {author} {\bibfnamefont {G.}~\bibnamefont
  {Brassard}},\ }\href@noop {} {}\bibinfo {howpublished} {Proceedings of the
  IEEE International Conference on Computers, Systems and Signal Processing,
  Bangalore, India,175} (\bibinfo {year} {1984})\BibitemShut {NoStop}%
\bibitem [{\citenamefont {Tanzilli}\ \emph {et~al.}(2012)\citenamefont
  {Tanzilli}, \citenamefont {Martin}, \citenamefont {Kaiser}, \citenamefont
  {De~Micheli}, \citenamefont {Alibart},\ and\ \citenamefont
  {Ostrowsky}}]{tanzilli2012}%
  \BibitemOpen
  \bibfield  {author} {\bibinfo {author} {\bibfnamefont {S.}~\bibnamefont
  {Tanzilli}}, \bibinfo {author} {\bibfnamefont {A.}~\bibnamefont {Martin}},
  \bibinfo {author} {\bibfnamefont {F.}~\bibnamefont {Kaiser}}, \bibinfo
  {author} {\bibfnamefont {M.}~\bibnamefont {De~Micheli}}, \bibinfo {author}
  {\bibfnamefont {O.}~\bibnamefont {Alibart}}, \ and\ \bibinfo {author}
  {\bibfnamefont {D.}~\bibnamefont {Ostrowsky}},\ }\href {\doibase
  10.1002/lpor.201100010} {\bibfield  {journal} {\bibinfo  {journal} {Laser \&
  Photonics Reviews}\ }\textbf {\bibinfo {volume} {6}},\ \bibinfo {pages} {115}
  (\bibinfo {year} {2012})}\BibitemShut {NoStop}%
\bibitem [{PSE()}]{PSEUDORNG}%
  \BibitemOpen
  \href
  {http://www.xilinx.com/support/documentation/application_notes/xapp052.pdf}
  {}\bibinfo {note} {Xilinx: Efficient Shift Registers, LFSR Counters, and Long
  Pseudo-Random Sequence Generators}\BibitemShut {NoStop}%
\bibitem [{\citenamefont {Wooten}\ \emph {et~al.}(2000)\citenamefont {Wooten},
  \citenamefont {Kissa}, \citenamefont {Yi-Yan}, \citenamefont {Murphy},
  \citenamefont {Lafaw}, \citenamefont {Hallemeier}, \citenamefont {Maack},
  \citenamefont {Attanasio}, \citenamefont {Fritz}, \citenamefont {McBrien},\
  and\ \citenamefont {Bossi}}]{Wooten2000}%
  \BibitemOpen
  \bibfield  {author} {\bibinfo {author} {\bibfnamefont {E.}~\bibnamefont
  {Wooten}}, \bibinfo {author} {\bibfnamefont {K.}~\bibnamefont {Kissa}},
  \bibinfo {author} {\bibfnamefont {A.}~\bibnamefont {Yi-Yan}}, \bibinfo
  {author} {\bibfnamefont {E.}~\bibnamefont {Murphy}}, \bibinfo {author}
  {\bibfnamefont {D.}~\bibnamefont {Lafaw}}, \bibinfo {author} {\bibfnamefont
  {P.}~\bibnamefont {Hallemeier}}, \bibinfo {author} {\bibfnamefont
  {D.}~\bibnamefont {Maack}}, \bibinfo {author} {\bibfnamefont
  {D.}~\bibnamefont {Attanasio}}, \bibinfo {author} {\bibfnamefont
  {D.}~\bibnamefont {Fritz}}, \bibinfo {author} {\bibfnamefont
  {G.}~\bibnamefont {McBrien}}, \ and\ \bibinfo {author} {\bibfnamefont
  {D.}~\bibnamefont {Bossi}},\ }\href {\doibase 10.1109/2944.826874} {\bibfield
   {journal} {\bibinfo  {journal} {Sel. Topics in Quantum Electr., IEEE Journal
  of}\ }\textbf {\bibinfo {volume} {6}},\ \bibinfo {pages} {69} (\bibinfo
  {year} {2000})}\BibitemShut {NoStop}%
\bibitem [{\citenamefont {Troyer}\ and\ \citenamefont
  {Renner}(2012)}]{troyer2012}%
  \BibitemOpen
  \bibfield  {author} {\bibinfo {author} {\bibfnamefont {M.}~\bibnamefont
  {Troyer}}\ and\ \bibinfo {author} {\bibfnamefont {R.}~\bibnamefont
  {Renner}},\ }\href
  {http://www.idquantique.com/images/stories/PDF/quantis-random-generator/quantis-rndextract-techpaper.pdf}
  {\enquote {\bibinfo {title} {A randomness extractor for the quantis
  device},}\ }\bibinfo {howpublished}
  {http://www.idquantique.com/images/stories/PDF/quantis-random-generator/quantis-rndextract-techpaper.pdf}
  (\bibinfo {year} {2012})\BibitemShut {NoStop}%
\bibitem [{\citenamefont {Mitchell}\ \emph {et~al.}(2015)\citenamefont
  {Mitchell}, \citenamefont {Abellan},\ and\ \citenamefont
  {Amaya}}]{mitchell2015}%
  \BibitemOpen
  \bibfield  {author} {\bibinfo {author} {\bibfnamefont {M.~W.}\ \bibnamefont
  {Mitchell}}, \bibinfo {author} {\bibfnamefont {C.}~\bibnamefont {Abellan}}, \
  and\ \bibinfo {author} {\bibfnamefont {W.}~\bibnamefont {Amaya}},\ }\href
  {\doibase 10.1103/PhysRevA.91.012314} {\bibfield  {journal} {\bibinfo
  {journal} {Phys. Rev. A}\ }\textbf {\bibinfo {volume} {91}},\ \bibinfo
  {pages} {012314} (\bibinfo {year} {2015})}\BibitemShut {NoStop}%
\bibitem [{\citenamefont {Ca{\~n}as}\ \emph {et~al.}(2014)\citenamefont
  {Ca{\~n}as} \emph {et~al.}}]{canas2014}%
  \BibitemOpen
  \bibfield  {author} {\bibinfo {author} {\bibfnamefont {G.}~\bibnamefont
  {Ca{\~n}as}} \emph {et~al.},\ }\href {http://arxiv.org/abs/1410.3443}
  {\bibfield  {journal} {\bibinfo  {journal} {arXiv e-print}\ } (\bibinfo
  {year} {2014})},\ \Eprint {http://arxiv.org/abs/1410.3443} {arXiv:1410.3443
  [quant-ph]} \BibitemShut {NoStop}%
\bibitem [{\citenamefont {Haw}\ \emph {et~al.}(2014)\citenamefont {Haw} \emph
  {et~al.}}]{haw2014}%
  \BibitemOpen
  \bibfield  {author} {\bibinfo {author} {\bibfnamefont {J.~Y.}\ \bibnamefont
  {Haw}} \emph {et~al.},\ }\href {http://arxiv.org/abs/1411.4512} {\bibfield
  {journal} {\bibinfo  {journal} {arXiv e-print}\ } (\bibinfo {year} {2014})},\
  \Eprint {http://arxiv.org/abs/1411.4512} {arXiv:1411.4512 [quant-ph]}
  \BibitemShut {NoStop}%
\bibitem [{\citenamefont {Hoeffding}(1963)}]{hoeffding1963}%
  \BibitemOpen
  \bibfield  {author} {\bibinfo {author} {\bibfnamefont {W.}~\bibnamefont
  {Hoeffding}},\ }\href {\doibase 10.1080/01621459.1963.10500830} {\bibfield
  {journal} {\bibinfo  {journal} {Journal of the American Statistical
  Association}\ }\textbf {\bibinfo {volume} {58}},\ \bibinfo {pages} {13}
  (\bibinfo {year} {1963})}\BibitemShut {NoStop}%
\end{thebibliography}

%

\end{document}